\documentclass[usenatbib,soprop,useAMS]{mn2e}
\usepackage{color}
\usepackage{mathptmx}
\usepackage{amssymb,amsmath}
\usepackage{graphicx}
\usepackage{aas_macros}
\usepackage[draft]{hyperref}

\newcommand{\teff}{\mathrm{T_{eff}}}
\newcommand{\logg}{\mathrm{\log~g}}
\newcommand{\meta}{\mathrm{[M/H]}}
\newcommand{\vphi}{\mathrm{V_\phi}}
\newcommand{\kms}{km~s$^{-1}$}

\newcommand{\vrdmcyl}{$-2 \pm 1$}
\newcommand{\vphidmcyl}{$215 \pm 1$}
\newcommand{\vzdmcyl}{$0 \pm 1$}
\newcommand{\metadmcyl}{$-0.09 \pm 0.01$}

\newcommand{\svrdmcyl}{$30 \pm 1$}
\newcommand{\svphidmcyl}{$20 \pm 1$}
\newcommand{\svzdmcyl}{$18 \pm 1$}
\newcommand{\smetadmcyl}{$0.18 \pm 0.01$}

\newcommand{\nvrdmcyl}{74\%}
\newcommand{\nvphidmcyl}{84\%}
\newcommand{\nvzdmcyl}{86\%}
\newcommand{\nmetadmcyl}{81\%}

\newcommand{\vrdecyl}{$ 2 \pm 1$}
\newcommand{\vphidecyl}{$180 \pm 9$}
\newcommand{\vzdecyl}{$-4 \pm 2$}
\newcommand{\metadecyl}{$-0.45 \pm 0.01$}

\newcommand{\svrdecyl}{$61 \pm 3$}
\newcommand{\svphidecyl}{$45 \pm 2$}
\newcommand{\svzdecyl}{$44 \pm 3$}
\newcommand{\smetadecyl}{$0.26 \pm 0.01$}

\newcommand{\nvrdecyl}{24\%}
\newcommand{\nvphidecyl}{15\%}
\newcommand{\nvzdecyl}{12\%}
\newcommand{\nmetadecyl}{18\%}

\newcommand{\vrhacyl}{$ 7 \pm 4$}
\newcommand{\vphihacyl}{$15 \pm 14$}
\newcommand{\vzhacyl}{$-7 \pm 3$}
\newcommand{\metahacyl}{$-1.25 \pm 0.15$}

\newcommand{\svrhacyl}{$160 \pm 19$}
\newcommand{\svphihacyl}{$119 \pm 19$}
\newcommand{\svzhacyl}{$110 \pm 12$}
\newcommand{\smetahacyl}{$0.56 \pm 0.05$}

\newcommand{\nvrhacyl}{2\%}
\newcommand{\nvphihacyl}{$1$\%}
\newcommand{\nvzhacyl}{2\%}
\newcommand{\nmetahacyl}{$1$\%}

\newcommand{\vrdmTD}{$0 \pm 3$}
\newcommand{\vphidmTD}{$209 \pm 3$}
\newcommand{\vzdmTD}{$-1\pm 1$}
\newcommand{\metadmTD}{$-0.15 \pm 0.01 $}

\newcommand{\svrdmTD}{$39 \pm 4$}
\newcommand{\svphidmTD}{$30 \pm 3$}
\newcommand{\svzdmTD}{$27 \pm 2$}
\newcommand{\smetadmTD}{$0.15 \pm 0.01$}

\newcommand{\nvrdmTD}{10\%}
\newcommand{\nvphidmTD}{ 13\%}
\newcommand{\nvzdmTD}{13\%}
\newcommand{\nmetadmTD}{4\%}

\newcommand{\vrdeTD}{$9 \pm 3$}
\newcommand{\vphideTD}{$177 \pm 2$}
\newcommand{\vzdeTD}{$-1 \pm 1$}
\newcommand{\metadeTD}{$-0.45 \pm 0.01$}

\newcommand{\svrdeTD}{$87 \pm 5$}
\newcommand{\svphideTD}{$60 \pm 1$}
\newcommand{\svzdeTD}{$55 \pm 2$}
\newcommand{\smetadeTD}{$0.21 \pm 0.01$}

\newcommand{\nvrdeTD}{79\%}
\newcommand{\nvphideTD}{81\%}
\newcommand{\nvzdeTD}{78\%}
\newcommand{\nmetadeTD}{89\%}

\newcommand{\vrhaTD}{$ 7 \pm 6$}
\newcommand{\vphihaTD}{$-1 \pm 11$}
\newcommand{\vzhaTD}{$-4 \pm 5$}
\newcommand{\metahaTD}{$-1.20 \pm 0.01$}

\newcommand{\svrhaTD}{$158 \pm  6$}
\newcommand{\svphihaTD}{$96 \pm 7$}
\newcommand{\svzhaTD}{$128 \pm 8$}
\newcommand{\smetahaTD}{$0.54 \pm 0.04$}

\newcommand{\nvrhaTD}{11\%}
\newcommand{\nvphihaTD}{6\%}
\newcommand{\nvzhaTD}{9\%}
\newcommand{\nmetahaTD}{6\%}

\newcommand{\vrdeMWTD}{$-7 \pm 4$}
\newcommand{\vphideMWTD}{$123 \pm 16$}
\newcommand{\vzdeMWTD}{$0 \pm 6$}

\newcommand{\svrdeMWTD}{$89 \pm 7$}
\newcommand{\svphideMWTD}{$91 \pm 13$}
\newcommand{\svzdeMWTD}{$81 \pm 7$}

\newcommand{\nvrdeMWTD}{46\%}
\newcommand{\nvphideMWTD}{59\%}
\newcommand{\nvzdeMWTD}{49\%}

\newcommand{\vrhaMWTD}{$ 5\pm 5$}
\newcommand{\vphihaMWTD}{$-4\pm 1$}
\newcommand{\vzhaMWTD}{$-4 \pm 9$}

\newcommand{\svrhaMWTD}{$159 \pm  2$}
\newcommand{\svphihaMWTD}{$108 \pm 4$}
\newcommand{\svzhaMWTD}{$117 \pm 14$}

\newcommand{\nvrhaMWTD}{54\%}
\newcommand{\nvphihaMWTD}{41\%}
\newcommand{\nvzhaMWTD}{51\%}

\defcitealias{Kordopatis13b}{K13}
\defcitealias{Binney13}{B13}

\title{In the thick of it: metal-poor disc stars in RAVE}

\author[G. Kordopatis et al.]
{G. Kordopatis\thanks{E-mail:~ gkordo@ast.cam.ac.uk},$^1$~
 G. Gilmore, $^1$~
R.F.G.~Wyse,$^2$~
M.~Steinmetz,$^3$~ 
A.~Siebert,$^4$~
O.~Bienaym\'e,$^4$~ \newauthor
P.J.~McMillan,$^5$~ 
I.~Minchev,$^3$~
T.~Zwitter,$^{6,7}$~
B.K.~Gibson,$^8$~ 
G.~Seabroke,$^{9}$~
E.K.~Grebel, $^{10}$~\newauthor
J. ~Bland-Hawthorn, $^{11}$~
C.~Boeche,$^{10}$~
K.C.~Freeman,$^{12}$~
U.~Munari,$^{13}$~
J.F.~Navarro,$^{14}$~ \newauthor
Q.~Parker,$^{15,16,17}$~
W.A.~Reid,$^{15,16}$~ 
A.~Siviero$^{3,18}$\\
$^1$ Institute of Astronomy, University of Cambridge, Madingley Road, Cambridge CB3 0HA, UK\\
$^2$ Johns Hopkins University, 3400 N Charles Street, Baltimore, MD, 21218, USA\\
$^3$ Leibniz-Institut f\"ur  Astrophysik Potsdam (AIP), An der Sternwarte 16, 14482 Potsdam, Germany\\
$^4$ Observatoire astronomique de Strasbourg, Universit\'e de Strasbourg, CNRS, UMR 7550, 11 rue de l'Universit\'e, F-67000 Strasbourg, France\\
$^5$ Rudolf Peierls Centre for Theoretical Physics, Keble Road, Oxford, OX1 3NP, UK \\
$^{6}$ {Faculty of Mathematics and Physics, University of Ljubljana, Jadranska 19, 1000 Ljubljana, Slovenia}\\
$^{7}$ {Center of Excellence SPACE-SI, Askerceva cesta 12, 1000 Ljubljana, Slovenia}\\
$^8$ Jeremiah Horrocks Institute, University of Central Lancashire, Preston, PR1 2HE, UK \\
$^{9}$ Mullard Space Science Laboratory, University College London, Holmbury St Mary, Dorking, RH5 6NT, UK\\
$^{10}$ {Astronomisches Rechen-Institut, Zentrum f\"ur Astronomie der Universit\"at Heidelberg, M\"onchhofstr.\ 12-14, D-69120 Heidelberg, Germany}\\
$^{11}$ Sydney Institute for Astronomy, University of Sydney A28, School of Physics, NSW 2006, Australia \\
$^{12}$~{Research School of Astronomy and Astrophysics, Australian National University, Cotter Rd., Weston, ACT 2611, Australia}\\
$^{13}$ INAF Astronomical Observatory of Padova, 36012 Asiago (VI), Italy \\
$^{14}$ CIfAR Senior Fellow, Department of Physics and Astronomy, University of Victoria, Victoria BC, Canada V8P 5C2\\
$^{15}$ Department of Physics and Astronomy, Macquarie University, Sydney, NSW 2109 Australia\\
$^{16}$ Research Centre for Astronomy, Astrophysics and Astrophotonics, Macquarie University, Sydney, NSW 2109 Australia \\
$^{17}$ Australian Astronomical Observatory, PO Box 915, North Ryde NSW 1670\\
$^{18}$ Department of Physics and Astronomy, Padova University, Vicolo dell'Osservatorio 2, I-35122 Padova, Italy 
}

\date{Accepted 2013 September 23. Received 2013 September 18; in original form 2013 July 29 }

\begin{document}

 \maketitle

\begin{abstract}

By selecting in the RAVE-DR4 survey the stars located between 1 and 2 kpc above the Galactic plane, we question the consistency of the simplest three-component model (thin disc, thick disc, halo) for the Milky Way. 
We confirm that the metallicity and azimuthal velocity distribution functions of the thick disc are not Gaussian. In particular, we find that the thick disc has an extended metallicity tail going at least down to $\meta=-2$~dex, contributing roughly at 3\% of the entire thick disc population and having a shorter scale-length compared to the canonical thick disc. The mean azimuthal velocity of these metal-poor stars allows us to estimate the correlation between the metallicity ($\meta$) and the orbital velocity ($\vphi$), which is an important constraint on the formation mechanisms of the Galactic thick disc. Given our simple approach, we find  $\partial \vphi  / \partial \meta \approx  50$~\kms~dex$^{-1}$, which is in a very good agreement with previous literature values.
We complete the study with a brief discussion on the implications of the formation scenarios for the thick disc, and suggest that given the above mentioned characteristics, a thick disc mainly formed by radial migration mechanisms seems unlikely.

\end{abstract}


\section{Introduction}

The Galactic models and simulations of the Milky Way often separate the old stellar populations in four chemically and kinematically distinctive components, namely the thin disc, the thick disc, the halo and the bulge.  Whereas it is rather well established that the thin disc \citep[e.g.:][]{Antoja11, Boeche13}, the halo \citep[e.g.:][]{Carollo10} and the bulge \citep[e.g.:][]{Zoccali08,Hill11} can be further decomposed into sub-populations, there is still a debate whether it is the case for the thick disc \citep{Gilmore02, Wyse06}, or even if the thick disc is indeed distinct from the thin disc \citep*{Bovy12}.

The different mechanisms invoked in order to explain the origin of the Galactic thick disc, imply that  this structure is closely related to the formation and the evolution of the Milky Way itself \citep{Rix13}. The existence of the thick disc was first suggested by \cite{Gilmore83}, and since then a lot of effort has been made in order to measure its general properties. In the Solar neighbourhood, the thick disc is thought to be composed of old \citep[10~Gyr,][]{Furmann08} and relatively metal-poor stars\footnote{Defined the usual way by $\meta=\log({\rm M / H})_\star-\log({\rm M / H})_\odot$, with M including all the elements heavier than He. } \citep[$\meta \sim -0.5$~dex,][]{Bensby07,Kordopatis11b}, enhanced in $\alpha-$elements compared to the thin disc stars \citep[e.g.: ][]{Mishenina04, Bensby05, Reddy06, Adibekyan13}. In addition, the stars belonging to the thick disc are dynamically hotter than the thin disc ones, with a vertical velocity dispersion twice as big as the thin disc and a lag in the azimuthal rotation, relative to the Local Standard of Rest (LSR)  between 30 and 50~\kms \citep[e.g.:][]{Lee11, Pasetto12}.

The absence until recently of a large statistical catalogue of stars with precise proper motions, distances, and atmospheric parameters (effective temperature, surface gravity, metallicity) prevented the scientific community from having important constraints concerning the tails of the distribution functions (DF) of the thick disc,  for both the velocities and the metallicity. However, this information is essential, since  the shape of the metallicity distribution function (MDF) would constrain the star formation history of the Galaxy \citep[e.g.:][]{Calura12,Pilkington12}. In particular, such information would allow to disentangle between the
different thick disc formation scenarios available in the literature \citep[e.g:][]{Abadi03, Brook04, Villalobos08, Bournaud09, Schonrich10, Loebman11, Minchev12}.

Previous studies trying to investigate the tails of the MDF were usually selecting their targets based on the stellar 3-dimensional velocities, in order to associate probabilistically the stars with the thick disc \citep[e.g.:][]{Bensby07,Ruchti11}. In this study, we will put no constraints on the kinematics in order to achieve that goal.
Using for the first time an extended catalogue of stars comprising $5 \times 10 ^5$ targets, the {\it RAdial Velocity Experiment} (RAVE)  survey \citep{Steinmetz06}, we investigate whether the Milky Way is consistent with being a mixture of  three stellar populations, namely the thin disc, the thick disc and the halo\footnote{The selection function of RAVE does not allow to reach distances large enough to observe the Galactic bulge.}.  
Based on the fourth data release (DR4) parameters \citep[][K13 hereafter]{Kordopatis13b}, and on the distances computed by \citet{Binney13} (B13 hereafter, submitted), we investigate the properties of this kinematically unbiased sample of stars by fitting Gaussians in the velocity space, which are a good \emph{first} approximation to model the observations. 

The paper is structured as follows: Section~\ref{sect:general_sample} presents the general sample of stars that has been considered in the present study, as well as the assumptions made in order to compute the velocities. In Section~\ref{sect:Solar_neighbourhood} we are using a test subsample of stars in order to verify that our Gaussian fitting procedure can recover the commonly admitted kinematic, metallicity and density parameters for the Milky Way.  Then, in Section~\ref{sect:metal_poor} we select the metal-poor stars of the survey and investigate their kinematic properties, in particular their link with the thick disc. Section~\ref{sect:discussion} discusses the implications of an extended metal-poor tail of the thick disc, in relation with the most commonly cited thick disc formation mechanisms. Finally, Section~\ref{sect:conclusions} sums up.

\section{Sample selection and computation of the velocities}
\label{sect:general_sample}

\begin{figure*}
\centering

$\begin{array}{cc}
\includegraphics[width=0.4\linewidth, angle=0]{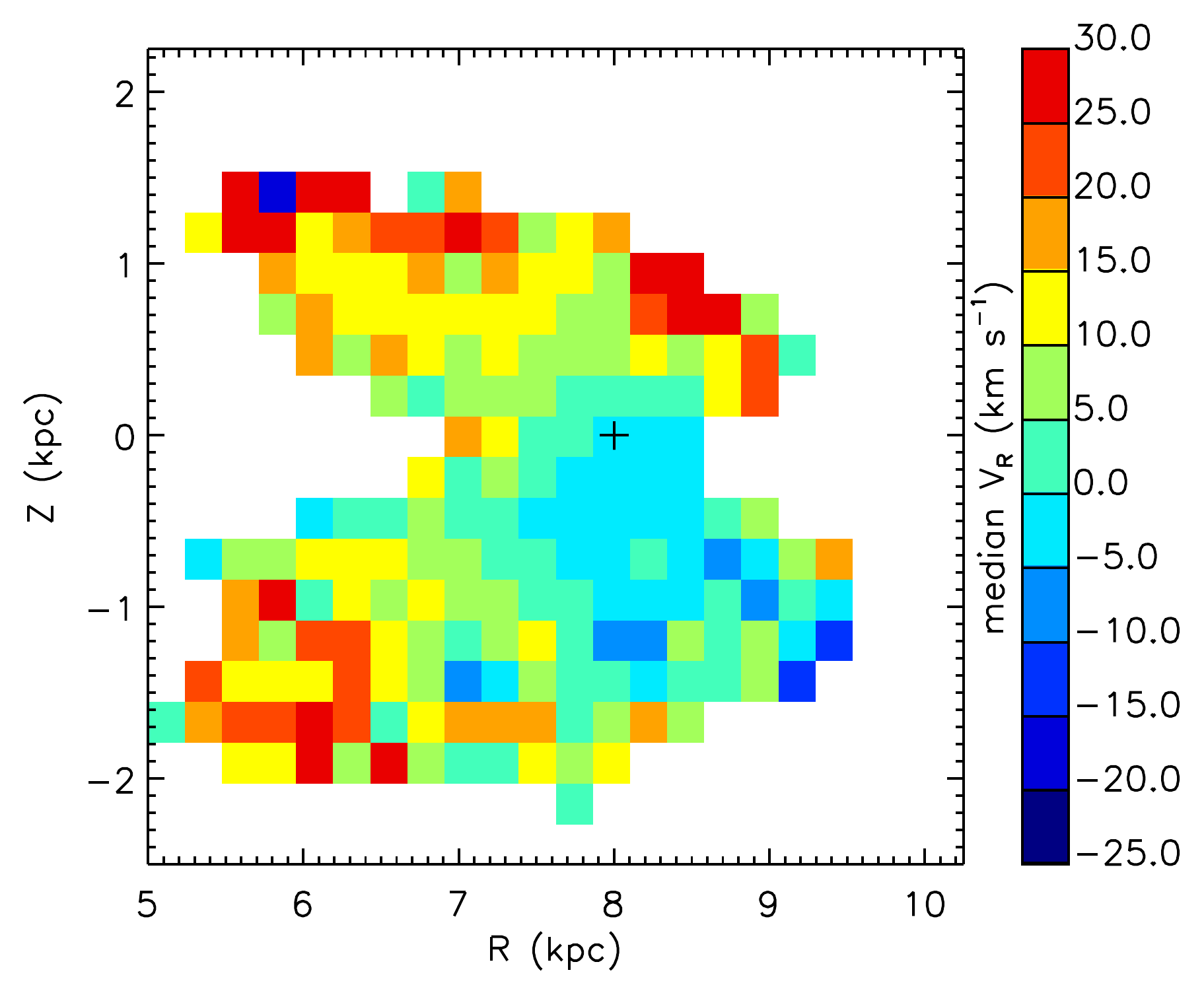}&
\includegraphics[width=0.4\linewidth, angle=0]{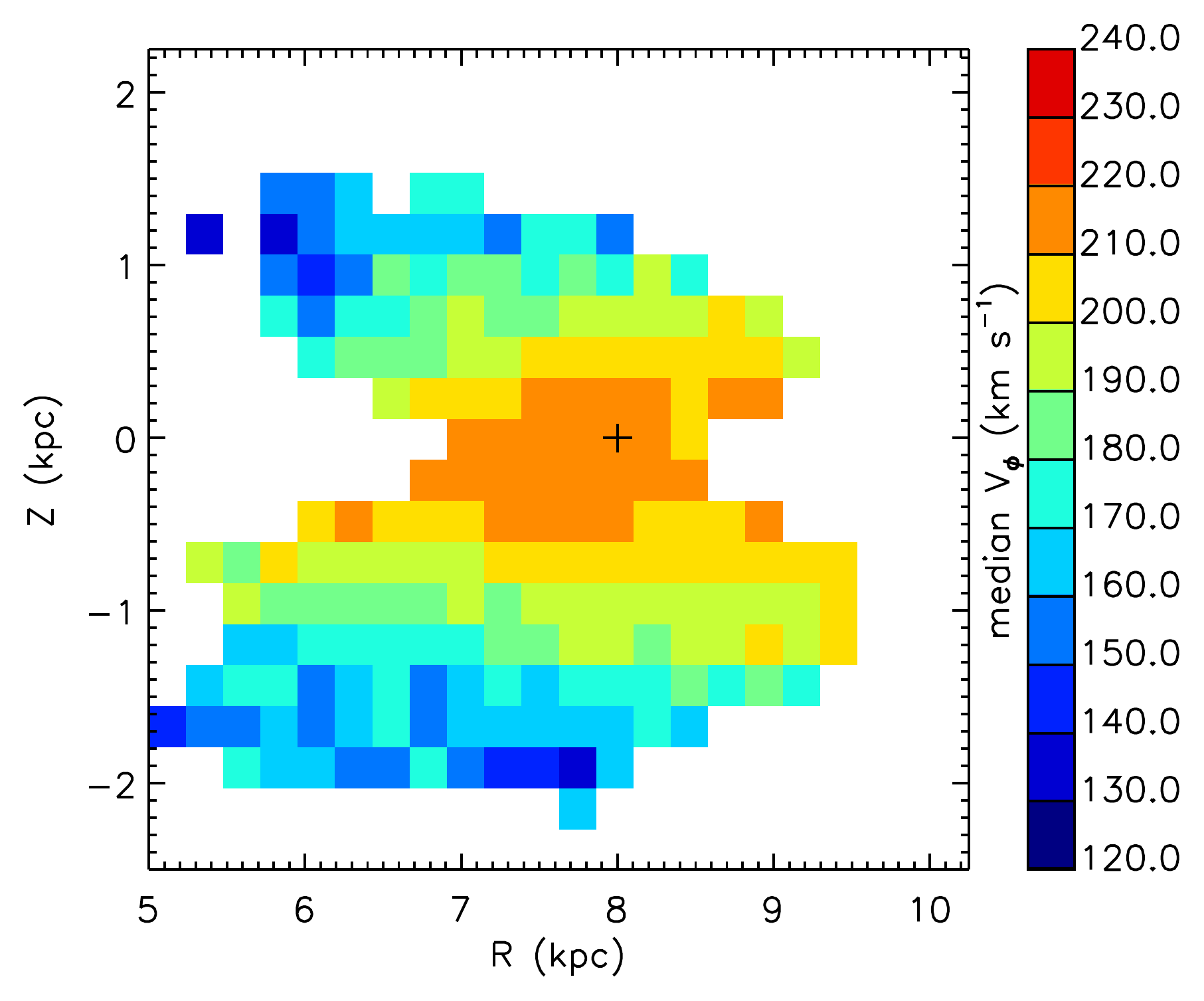}\\
\includegraphics[width=0.4\linewidth, angle=-0]{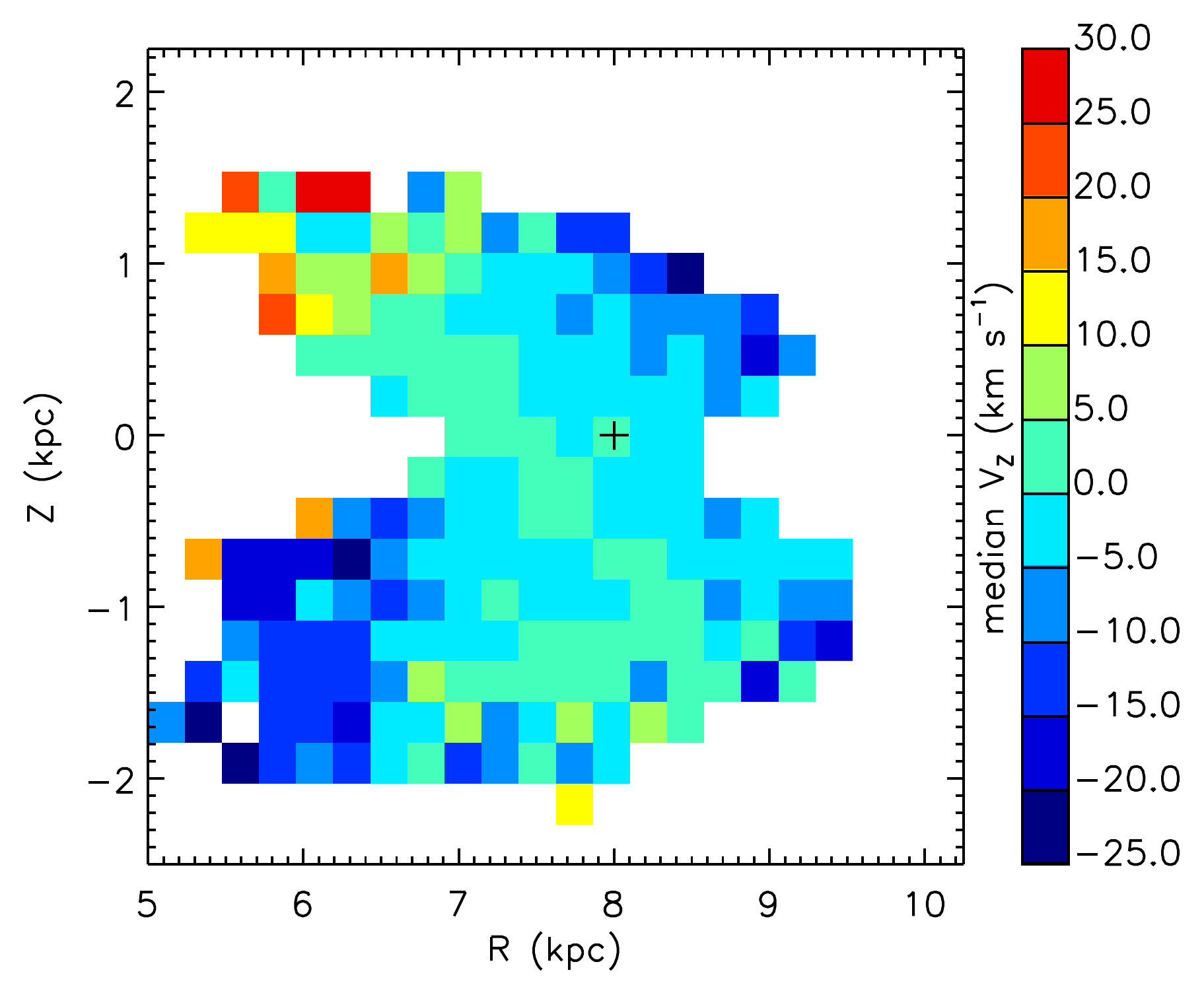}&
\includegraphics[width=0.4\linewidth, angle=0]{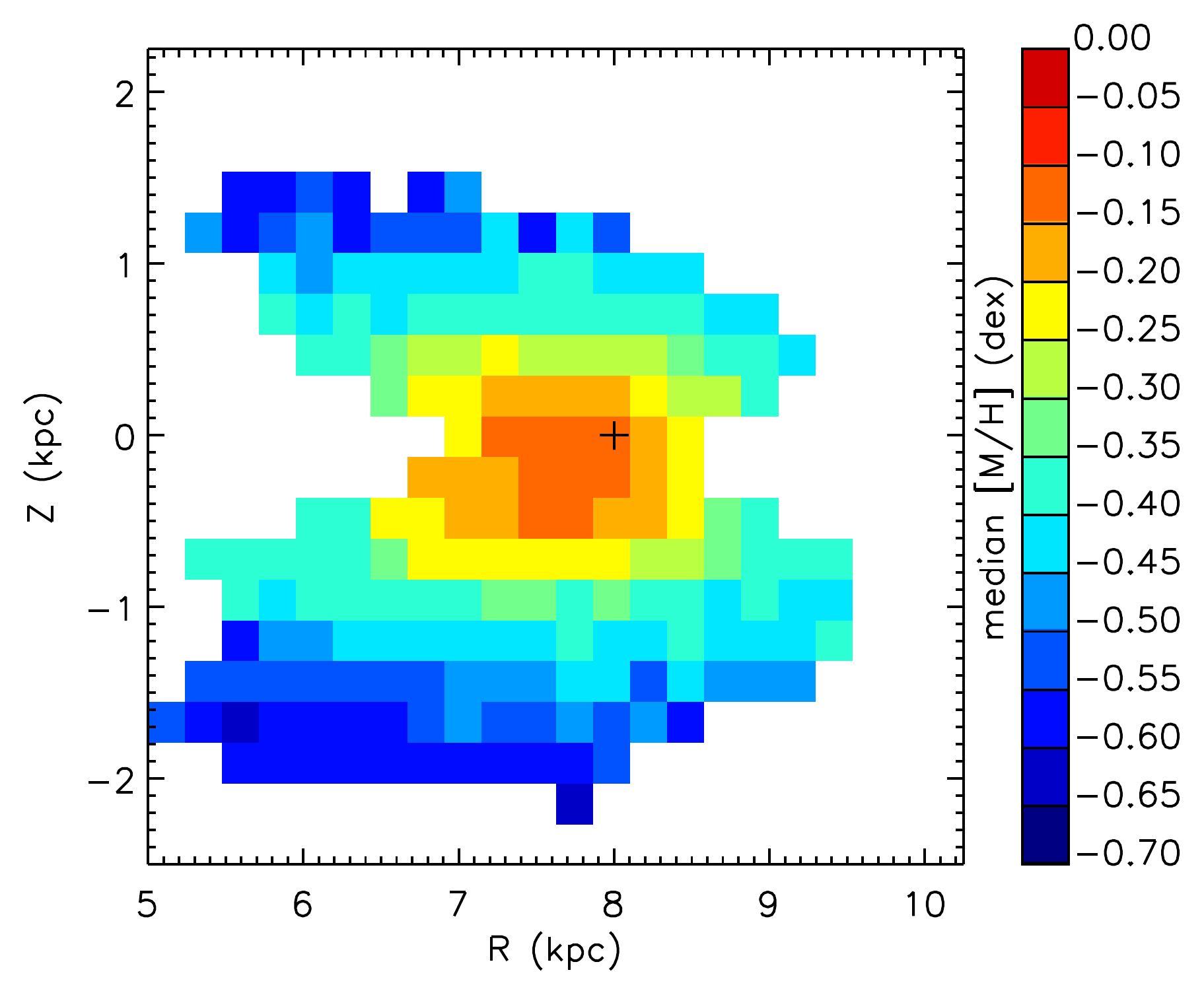}
\end{array}$

\caption{Median velocities and metallicities for all the RAVE stars for which distance and velocity determinations were available and fulfilling the selection criteria presented in Sect.~\ref{sect:general_sample}. The bins in $R$ and $Z$ are 0.25~kpc and contain at least 50 stars each. The black $"+"$ sign is at ($R_\odot=8$~kpc, $Z_\odot=0$~kpc), the assumed position of the Sun throughout this study.}
\label{fig:Overview}
\end{figure*}

The sample we considered for this study consists of the internal DR4 release, for which proper motions are a compilation of the PPMXL \citep{Roeser10}, Tycho-2 \citep{Hog00}, UCAC2 \citep{Zacharias04}, UCAC3 \citep{Zacharias10}, UCAC4 \citep{Zacharias13} and SPM4 sources \citep{Girard11}. In addition, the distances, parallaxes and extinctions have been computed using the \citetalias{Kordopatis13b} atmospheric parameters, and the Bayesian distance finding method presented in B13. This sample includes only stars with spectra for which the signal-to-noise ratio (SNR) is better than 20~pixel$^{-1}$, observed before 2012. In particular, it does not include multiple entries, since when different spectra were available for the same target, the B13 distances have been computed only for the one having the highest SNR.

We removed from our analysis the spectra for which the heliocentric radial velocity uncertainties were larger than 8~\kms, 
as well as the ones whose first three morphological flags do not indicate that they are normal stars
\citep[see][for details on how these flags are computed]{Matijevic12}. Indeed, for non-normal single stars at the rest-frame, the derived atmospheric parameters (and hence distances) obtained by the spectral analysis of \citetalias{Kordopatis13b} are uncertain. Furthermore, we removed the targets at low Galactic latitudes ($|b|<10^\circ,\sim 9\%$~of the stars) because the stronger inter-stellar extinction at these latitudes might bias the estimations of  both the atmospheric parameters\footnote{\citetalias{Kordopatis13b} uses the reddened 2MASS $(J-K_s)$ colour in order to constrain the spectral degeneracies by restricting the parameter range in the solution space. } and the derived distances.
In addition, in order to have reliable orbital velocities we removed from our sample all the stars that had estimated errors in the proper motions greater than 7~mas~year$^{-1}$ ($\sim$0.3\% of the stars).
Finally, it should be noted that, as recommended by B13, the distance estimator that is used in this study is the parallax ($\varpi$) and not the inferred distance itself, because of its higher reliability. Again, we removed the targets for which we had the worse parallax estimation. We arbitrarily set this threshold to $\sigma_\varpi/\varpi=0.8$, removing 
7\%  of the sample. It should be noted that lower $\sigma_\varpi/\varpi$ thresholds have been considered (e.g.: $\sigma_\varpi/\varpi< 0.5$), however without changing significantly the results of this study, since all the analysis is done by averaging the results from 500 Monte-Carlo realisations which are taking into account the individual errors (see discussion below in this section). This rather high value $\sigma_\varpi/\varpi$ has thus been adopted as a compromise between star counts and data quality.

The 3-dimensional Galactocentric orbital velocities (radial component $V_R$, azimuthal component $\vphi$ and vertical component $V_Z$) have been computed in the cylindrical reference frame, using the equations presented in the appendix of \cite{Williams13}.
We assumed  $R_\odot = 8$~kpc, $V_{LSR}=220$~\kms~ and adopted the Sun's peculiar velocities of \cite*{Schonrich10}, $(U_\odot,V_\odot,W_\odot)=(11.10, 12.24, 7.25)$~\kms.
{ 
We note that the choice of these values rather than others \citep[e.g.:][$U_\odot= 14$~\kms, $V_{LSR}=238$~\kms]{Schonrich12} has no particular impact on the derived conclusions of this paper other than mostly a zero-order shift. As a matter of fact, different peculiar velocities or $V_{LSR}$ only have an influence for the closest stars (of the order of the adopted difference in the velocities), whereas for the farthest ones random errors on the distances and the proper motions are dominating.  
}

In order to estimate the errors in the velocities, the Galactocentric positions (radial, $R$, and vertical, $Z$)  and the distribution functions, we  propagated the individual errors independently on the parallaxes, the heliocentric radial velocities and the proper motions by performing 500 Monte-Carlo realisations.  In addition, because of the correlation between the atmospheric parameters, and specifically between the surface gravity and the metallicity \citepalias[see][]{Kordopatis13b}, we imposed an anti-correlation between the variations of the line-of-sight distances and the metallicity, in the sense that if the Monte-Carlo realisation was randomly increasing the distance, then the metallicity was randomly decreased.
This allowed us to obtain error estimations for each velocity measurement, by considering the $1\sigma$-dispersion of the realisations for a single star. Then we excluded from the final sample the few  stars that had errors larger than  500~\kms~ in any velocity component, or $|\vphi| > 800$~\kms.

Figure~\ref{fig:Overview} shows maps of the median kinematic and metallicity properties in the $R-Z$~plane, of the 
considered stars in the RAVE survey that are used in the next sections.
From the plots on this figure, one can see on one hand that the median azimuthal velocity and the metallicity decrease with distance from the Galactic plane, unlike $V_Z$ or $V_R$ which absolute values tend to increase with $Z$.
In addition, evidence of  the shallow radial gradient in $V_R$ reported in \citet{Siebert11} as well as the compression-rarefaction pattern in $V_Z$ identified in \citet{Williams13} using only Red Clump giants are also observed in our sample. However, these trends are considered to secondary in our analysis, and will not be discussed in what follows.

 \begin{figure}
\centering
$\begin{array}{c}
\includegraphics[width=0.8\linewidth, angle=0]{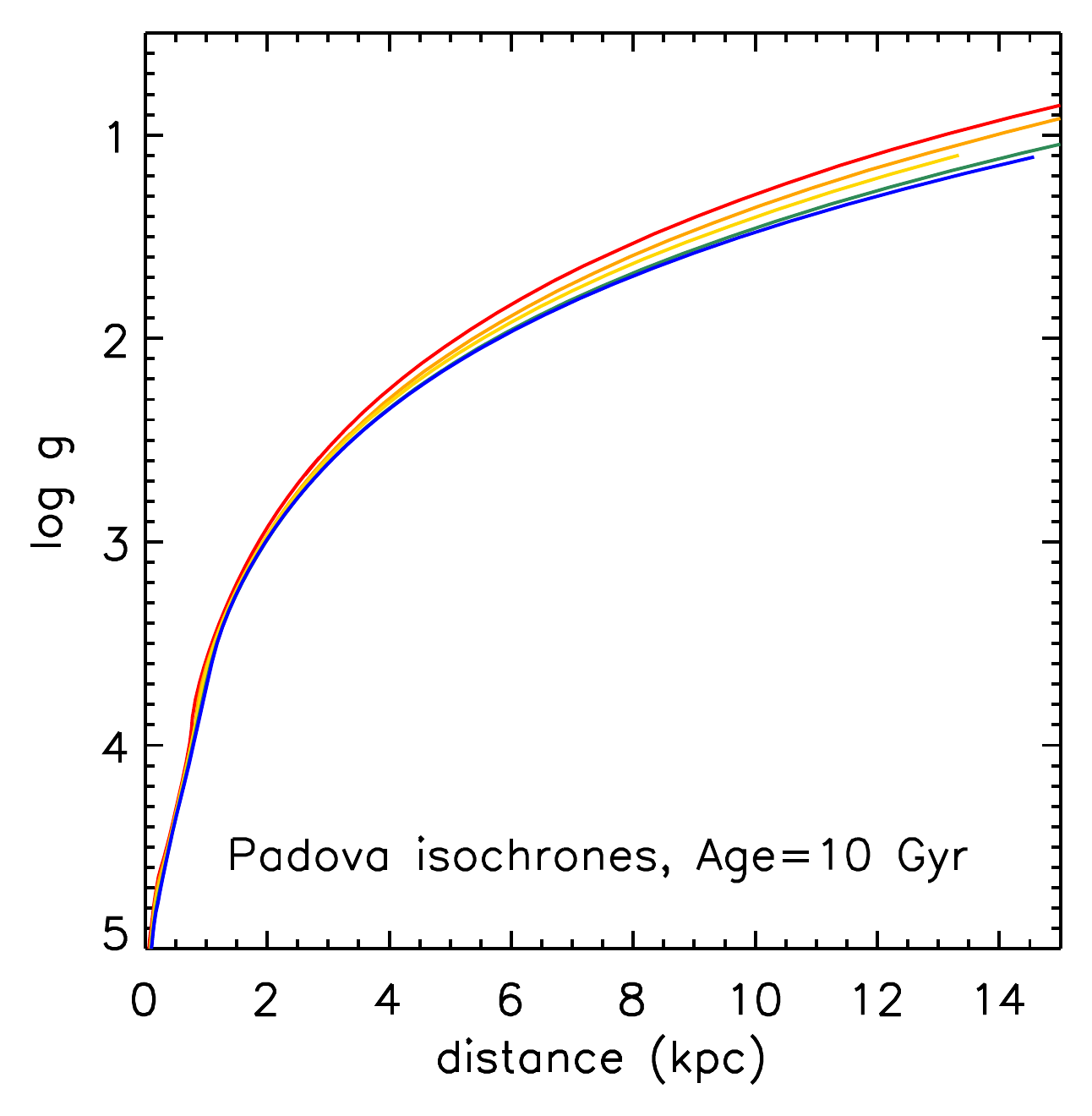} 
\end{array}$
\caption{Maximum observable line-of-sight distance as a function of surface gravity for stars with age=10~Gyr and for metallicity values of 0.0~dex (red curve), $-0.5$~dex (orange), $-1.0$~dex (yellow), $-1.5$~dex (green)  and $-2.0$~dex (blue), respectively. The considered isochrones are the ones of the Padova group, and the assumed limiting magnitude is $J=12$~mag.}
\label{fig:Malmquist_study}
\end{figure}

Finally, let us note that our analysis is done in a volume limited fashion, whereas RAVE is a magnitude complete sample. As a consequence, due to the fact that at a given colour (or effective temperature) metal-rich dwarf stars are intrinsically brighter than metal-poor dwarfs (and {\it vice versa} for the giants), biases might be introduced in our work. 
We used the Padova isochrones \citep{Bressan12} to investigate this effect. Figure~\ref{fig:Malmquist_study} shows the limiting line-of-sight distance at the RAVE limiting apparent magnitude of $J=12$~mag \citepalias[see][]{Kordopatis13b}, as a function of surface gravity and metallicity. 
For line-of-sight distances closer than 2~kpc ({\it i.e.} roughly the Solar neighbourhood defined in Sect.~\ref{sect:Solar_cylinder}), no significant biases are introduced, whereas for the farthest distances, metal-poor stars might be over-represented. Nevertheless, as it can be seen from Fig.~\ref{fig:Malmquist_study}, this effect should not alter significantly our analysis results.

\section{Characterisation of the RAVE sample in the Solar cylinder}
\label{sect:Solar_neighbourhood}

\subsection{The Solar neighbourhood}
\label{sect:Solar_cylinder}

\begin{figure}
\centering
$\begin{array}{c}
\includegraphics[width=0.9\linewidth, angle=0]{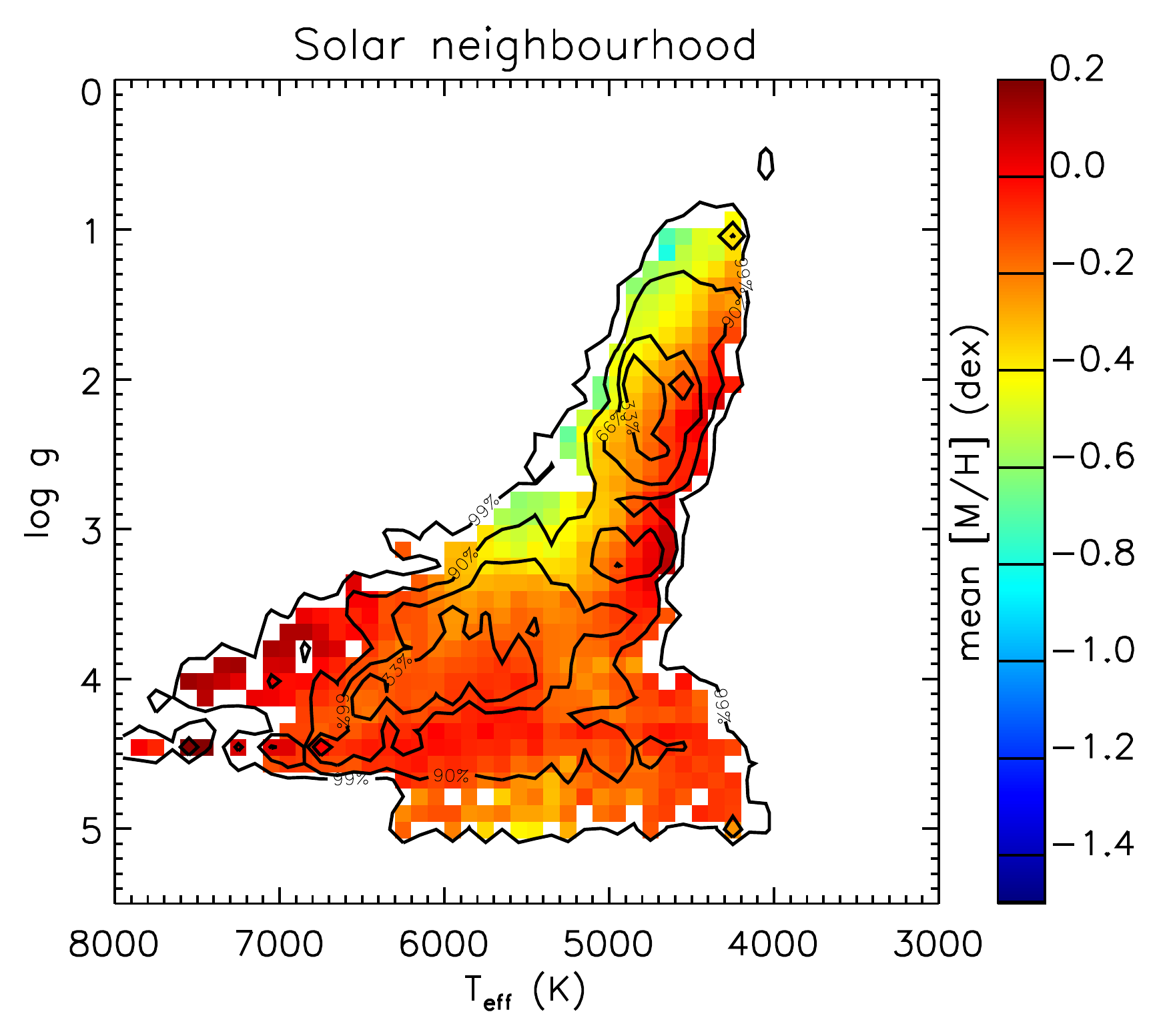}
\end{array}$
\caption{Mean metallicity across the Hertzsprung--Russell diagram of the selected stars in the Solar neighbourhood (see Sect.~\ref{sect:Solar_cylinder}).  The iso-contour lines contain 33\%, 66\%, 90\% and 99\% of the total selected sample. No particular selection biases are observed in our selection. }
\label{fig:HR_solar_neighbourhood}
\end{figure}

As a start, we select the stars that are in the Solar cylinder ($7.5 < R < 8.5$~kpc, 132012 stars) 
to try and model the metallicity and the velocity distributions. Therefore this sample  includes the stars that have been the most studied in past spectroscopic and photometric surveys \citep[e.g:][]{Nordstrom04,Juric08}, and it constitutes a volume inside which the properties of the Galactic components are relatively well established.
 Figure~\ref{fig:HR_solar_neighbourhood} shows the Hertzsprung--Russell (HR) diagram of the selected stars, colour-coded with the mean metallicity  of the targets at different combinations of effective temperature and surface gravity. One can see from this figure that the local sample does not suffer from any particular biases in metallicity (we find the same mean metallicity at all surface gravities), as already discussed in Sect.~\ref{sect:general_sample}.

 {  
Using a Truncated-Newton method \citep{Dembo83}, we maximised for each Monte-Carlo realisation the likelihood functions, $\mathcal{L}_\theta$, of the un-binned distributions of a given $\theta$ parameter ($\theta \equiv V_R, V_\phi, V_Z, \meta$), defined as: 
  \begin{equation}
   \mathcal{L}_\theta(\mbox{\footnotesize $ \rho_{1}, \mu_{1}, \sigma_{1}, \rho_{2}, \mu_{2}, \sigma_{2},\rho_{3}, \mu_{3}, \sigma_{3}$})=\\
    \prod_{i=1}^{N} \sum_{t=1}^{3}  \frac{\rho_{t}}{\sigma_{t}\sqrt{2\pi}} \cdot e^{  \mbox{\normalsize $-\frac{ (\theta_i-\mu_{t})^2 }  { 2 \sigma_{t}^2}$  }}
 \end{equation}
where $t$ is one of the Galactic populations (thin disc, thick disc, halo), $\rho$ is the relative weight of each population ($\rho_1+\rho_2+\rho_3=1$), $\mu$ and $\sigma$ are the means and dispersions of the Gaussian distributions, $i$ a given measurement of the $\theta$ parameter and $N$ the total number of stars. 

Due to the high number of parameters that we want to fit, initial conditions for the maximisation method have to be given.  These have been taken from our {\it a priori} expectation of the stellar populations characteristics in the Solar neighbourhood as commonly found in the more simplistic Galactic models in the literature.}
 In particular, the means ($\mu$), dispersions ($\sigma$) and normalisations of the Gaussian priors for each population  are:

\begin{itemize}

\item
For the (old) thin disc:  $(\mu_\meta,\sigma_{\meta}) \approx (- 0.1,0.2)$~dex, $(\mu_\vphi,\sigma_{\vphi}) \approx (210,25)$~\kms,  $(\mu_{V_R},\sigma_{V_R}) \approx (0,\sqrt{2} \sigma_\vphi)$~\kms, $(\mu_{V_Z},\sigma_{V_Z}) \approx (0,\sigma_\vphi)$~\kms\ \citep[e.g.: ][]{Robin03}.
We suppose that this population is the one dominating the sample, with roughly 85\% of the total number of stars \citep[e.g.:][]{Juric08}.

\item
For the canonical thick disc:  $(\mu_\meta,\sigma_{\meta}) \approx (- 0.5, 0.3)$~dex and $(\mu_\vphi,\sigma_{\vphi})=(180,50)$~\kms, $(\mu_{V_R},\sigma_{V_R}) \approx (0,\sqrt{2} \sigma_\vphi)$~\kms, $(\mu_{V_Z},\sigma_{V_Z}) \approx (0,\sigma_\vphi)$~\kms. The initial guess for the density of the thick disc is 10\% of the selected sample \citep[e.g.: ][]{Soubiran03}.

\item
For the stellar halo: $(\mu_\meta,\sigma_{\meta}) \approx (- 1.5, 0.5)$~dex, $(\mu_\vphi,\sigma_{\vphi})=(0,100)$~\kms, $(\mu_{V_R},\sigma_{V_R}) \approx (0,\sqrt{2} \sigma_\vphi)$~\kms, $(\mu_{V_Z},\sigma_{V_Z}) \approx (0,\sigma_\vphi)$~\kms\ \citep[e.g.: ][]{Chiba00}.

\end{itemize}

As it can be seen from the above priors, for all the stellar populations $\sigma_{V_R} / \sqrt{2} \approx \sigma_{\vphi} \approx \sigma_{V_Z}$  \citep[e.g.:][and references therein, and also \citet{Binney12} for possible deviations from this relation for thick disc stars]{Rix13}. 
 The most likely model is found by letting the mean velocities to vary by $\pm30$~\kms\ from the above mentioned mean values, and letting the dispersions vary by $\pm 10, 20, 30$~\kms\ for the thin, thick and halo, respectively. The mean metallicities are left to vary by $\pm0.1$~dex  for the thin and thick disc, and $\pm~0.3$~dex for the halo and the most likely dispersions are explored by varying by 0.1~dex their above mentioned metallicity dispersions. Finally, the relative density of the populations are left to vary freely.

 { 
We note that by averaging over 500 Monte-Carlo realisations we convolve the data with the errors for a second time, since  the raw velocities and metallicities include already measurement errors. This leads to distributions that are artificially broader than in reality, however without affecting their means (as long as the true error measurement distribution is close to Gaussian).  
For a given stellar population, the corrected dispersion, $\sigma_\theta$,  associated to a Gaussian distribution of a $\theta$ parameter, is obtained by quadratically subtracting the average measurement error $e_\theta$ of the stars, as follows:

\begin{equation}
\sigma_\theta=\sqrt{\sigma_{\theta,0}^2-2\cdot e_\theta^2},
\label{eqn:dispersion_correction}
\end{equation}
where $\sigma_{\theta,0}$ is the derived velocity dispersion of the parameter $\theta$ and the factor $2$ comes from the fact that the errors are accounted twice. Nevertheless, because we do not assign each star to one of the Galactic components (and thus not have an estimation of $e_\theta$) the  corrected dispersions of the Galactic components will only be derived and discussed in Sect.~\ref{sect:scale_lenghts}.
 }

\footnotesize

\begin{table*}
{ 
\caption{Means, dispersions and normalisations of the measured distributions functions for the Solar neighbourhood sample}}
\centering

\begin{tabular}{c c c c c c c c c c c c c}
\hline\hline
Galactic         & $V_R$ & $\vphi$ & $V_Z$ & $\sigma_{V_R}$& $\sigma_{V_\phi}$  & $\sigma_{V_Z}$ & $\meta$ & $\sigma_{\meta}$&  $N_{V_R}$ & $N_{\vphi}$ & $N_{V_Z}$ & $N_\meta$ \\
component        & \kms         & \kms            & \kms  & \kms         & \kms             & \kms & dex  & dex & & & \\
\hline
Thin disc & \vrdmcyl & \vphidmcyl & \vzdmcyl & \svrdmcyl & \svphidmcyl & \svzdmcyl & \metadmcyl & \smetadmcyl & \nvrdmcyl & \nvphidmcyl &\nvzdmcyl &\nmetadmcyl\\
Thick disc &\vrdecyl & \vphidecyl & \vzdecyl & \svrdecyl & \svphidecyl & \svzdecyl & \metadecyl & \smetadecyl & \nvrdecyl & \nvphidecyl &\nvzdecyl &\nmetadecyl\\
Halo       &\vrhacyl & \vphihacyl & \vzhacyl & \svrhacyl & \svphihacyl & \svzhacyl & \metahacyl & \smetahacyl & \nvrhacyl & \nvphihacyl &\nvzhacyl &\nmetahacyl  \\
\hline

\end{tabular}
\label{table:measured_vels_sol_cyl}
\end{table*}

\normalsize

\begin{figure}
\centering
$\begin{array}{c}
\includegraphics[width=1.0\linewidth, angle=0]{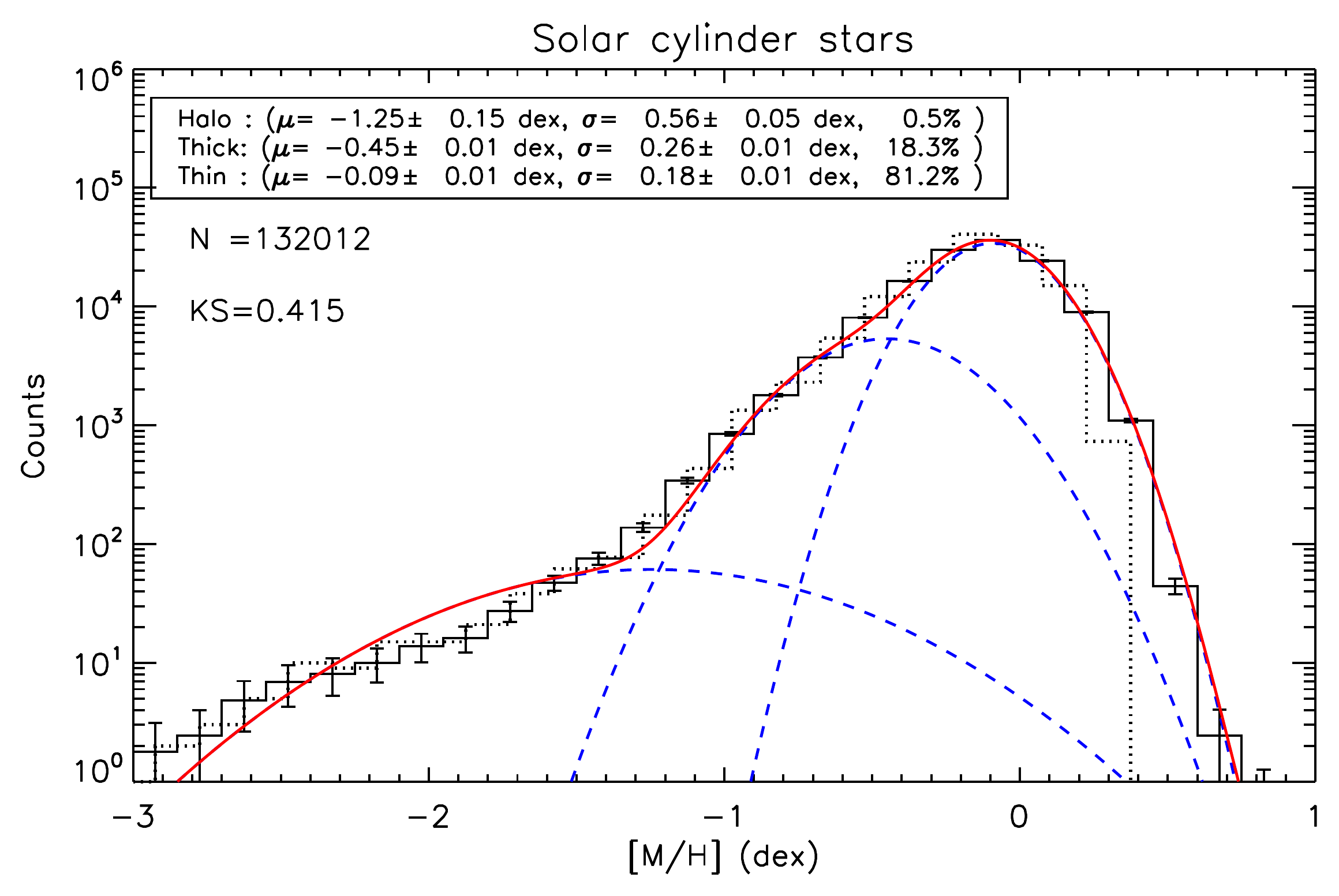} \\
\includegraphics[width=1.0\linewidth, angle=0]{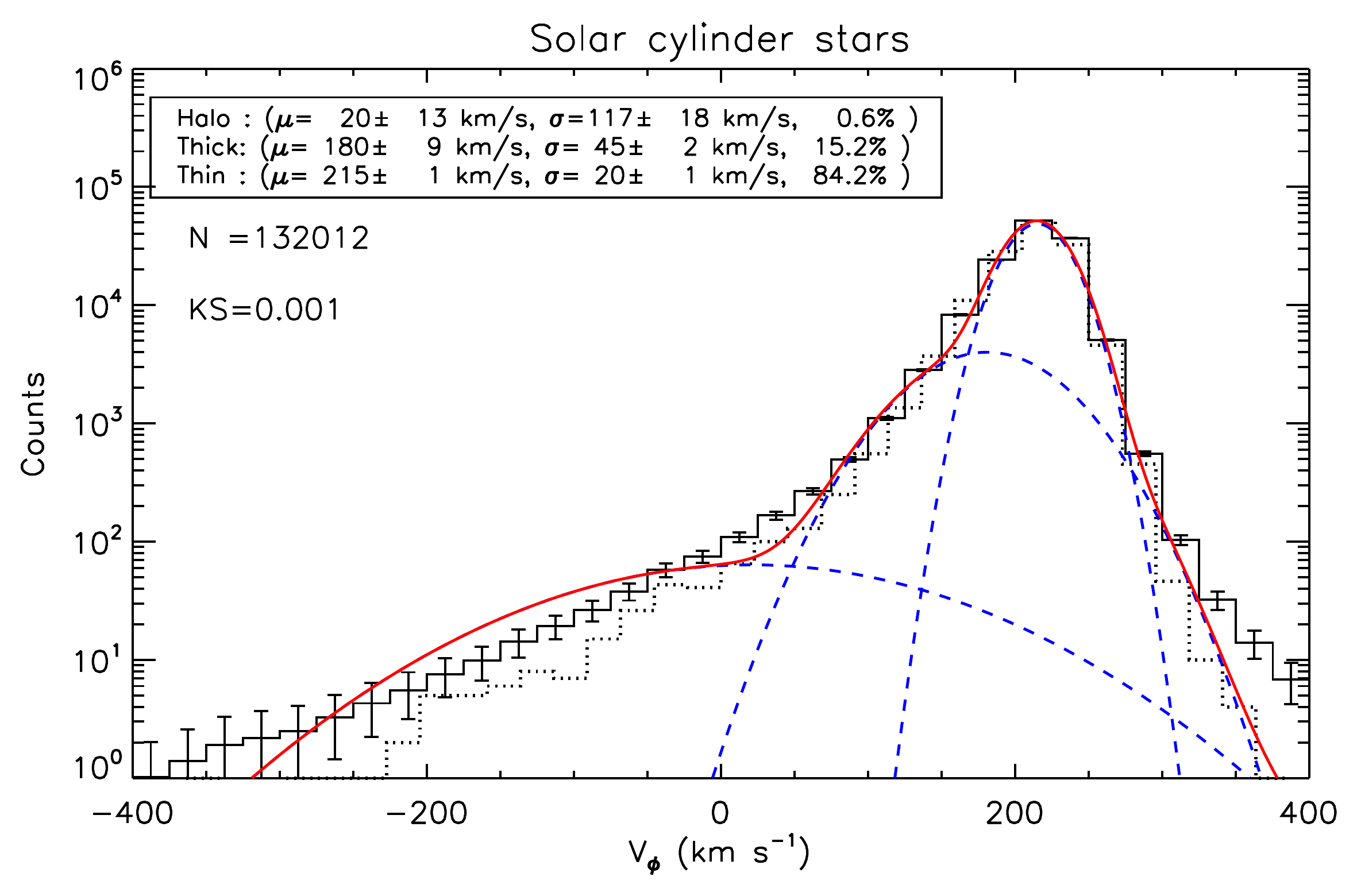}
\end{array}$
\caption{Probability distribution functions in a logarithmic scale for all the stars in the Solar neighbourhood, for the metallicities (upper panel, bin size=0.15~dex) and the azimuthal velocity (lower panel, bin size=30~\kms). Dotted-line histograms represent the actual data, whereas the plain-line histogram is averaged over 500 Monte-Carlo realisations. The error bars in the histograms correspond to Poisson noise ($\sqrt{N}$ in each bin). The individual Gaussians, corresponding to the thin disc, thick disc and stellar halo,  derived from our Maximum-Likelihood procedure are represented in dashed blue lines. Their means ($\mu$), dispersions ($\sigma$) and relative weight (in \%) are indicated in the upper left box. The sum of the Gaussians is plotted as a solid red line. Finally, the Kolmogorov-Smirnov probability of compatibility between the averaged observations and the averaged model is also indicated.}
\label{fig:Sol_cylinder}
\end{figure}

Table~\ref{table:measured_vels_sol_cyl} and Fig.~\ref{fig:Sol_cylinder} summarise and illustrate the means, the dispersions and the normalisation values of each population and for each distribution function, once averaged over the 500 Monte-Carlo realisations.
Though the fits are not perfect (due to the fact that the distribution functions are not truly Gaussians in reality, see for example Sharma et al., submitted), the Kolmogorov-Smirnov probabilities (noted KS-test hereafter),  measured in the ranges $-3 < $[M/H] $< +0.2$~dex, $-300< \vphi < +400$~kms (see discussion below),  are fairly consistent with the null hypothesis. 
In particular, we note that we find a thick disc lagging the LSR by $\sim 40$~\kms, 
{  and a slightly prograde halo,  most likely affected by the non-Gaussian velocity distribution tail of the discs extending to low $\vphi$ values \citep*[see for example][]{Schonrich11}.}

Depending on which distribution we fit, the total ratio of thin disc stars is between 74\% and 86\%, the one of the thick disc between 12\% and 24\%, and for the halo between 1\% and 2\%. Furthermore, these results are in agreement with the measured values of \citet{Williams13}, obtained using only the red clump giant stars of the sample (see their Fig.~14). 
We recall that the individual measurement errors on the velocities or the metallicities  have not been subtracted from the found fits, which results in systematically larger dispersions than the values found in reality ($i.e$ values larger by $\sim 0.05 - 0.1$~dex for the metallicities and $\sim 5 -10$~\kms\ for the velocities, see Sect.~\ref{sect:scale_lenghts}).
In addition, we highlight that not necessarily the same stars are associated with the different Galactic populations when considering the different velocity components or the metallicity. Nevertheless this is not an issue in this work, as we do not wish to assign a membership probability to each star, but rather investigate the shape of the distribution functions that best fit the data.

Furthermore, the adopted Gaussians over-predict the star counts above $\meta>+0.2$~dex. Nevertheless, it is difficult and beyond the scope of this paper  to state if this over-prediction is true or due to the calibration relation adopted in \citetalias{Kordopatis13b}. Indeed, we remind that the RAVE-DR4 metallicities are skewed at super-Solar abundances. This is due to the fact that very few reliable data sets are available in order to calibrate the high metallicity end of the RAVE-DR4 parameters (the calibrating sample with the highest metallicity corresponds to 6 spectra of the open cluster IC~4651, at [M/H]$\sim +0.10$~dex, see K13).

Finally, we also note that we fail to reproduce the distribution function for the stars having $\vphi \gtrsim 350$~\kms. These stars, for which we find unusually high velocities, are most likely a result of over-predictions of the inferred distances (e.g.: un-detected binaries affecting both the apparent magnitude and the radial velocity of the stars, or wrong extinction model to compute the distances). Indeed, tests have shown that decreasing the adopted distances by 15\% was affecting that high velocity tail, by reducing the number of these stars and hence achieving an agreement, within the error bars,  between the model Gaussians and the observations.

\subsection{The Galactic thick disc in the Solar cylinder}
\label{sect:thick_disc_Solar_neighbourhood}

\begin{table*}
{ 
\caption{Means, dispersions and normalisations of the measured distributions functions for the Solar cylinder sample at $1<|Z|<2$~kpc}}
\centering
\begin{tabular}{c c c c c c c c c c c c c}

\hline\hline
Galactic         & $V_R$ & $\vphi$ & $V_Z$ & $\sigma_{V_R}$& $\sigma_{V_\phi}$  & $\sigma_{V_Z}$ & $\meta$ & $\sigma_{\meta}$&  $N_{V_R}$ & $N_{\vphi}$ & $N_{V_Z}$ & $N_\meta$ \\
component        & \kms         & \kms             & \kms  & \kms         & \kms             & \kms & dex  & dex & & & \\
\hline
Thin disc & \vrdmTD & \vphidmTD & \vzdmTD & \svrdmTD & \svphidmTD & \svzdmTD & \metadmTD & \smetadmTD & \nvrdmTD & \nvphidmTD &\nvzdmTD  &\nmetadmTD\\
Thick disc &\vrdeTD & \vphideTD & \vzdeTD & \svrdeTD & \svphideTD & \svzdeTD & \metadeTD & \smetadeTD & \nvrdeTD & \nvphideTD &\nvzdeTD  &\nmetadeTD\\
Halo       &\vrhaTD & \vphihaTD & \vzhaTD & \svrhaTD & \svphihaTD & \svzhaTD & \metahaTD & \smetahaTD & \nvrhaTD & \nvphihaTD &\nvzhaTD  &\nmetahaTD  \\
\hline

\end{tabular}
\label{table:measured_vels_sol_cyl_thick_disc}
\end{table*}

\begin{figure}
\centering
$\begin{array}{c}
\includegraphics[width=0.9\linewidth, angle=0]{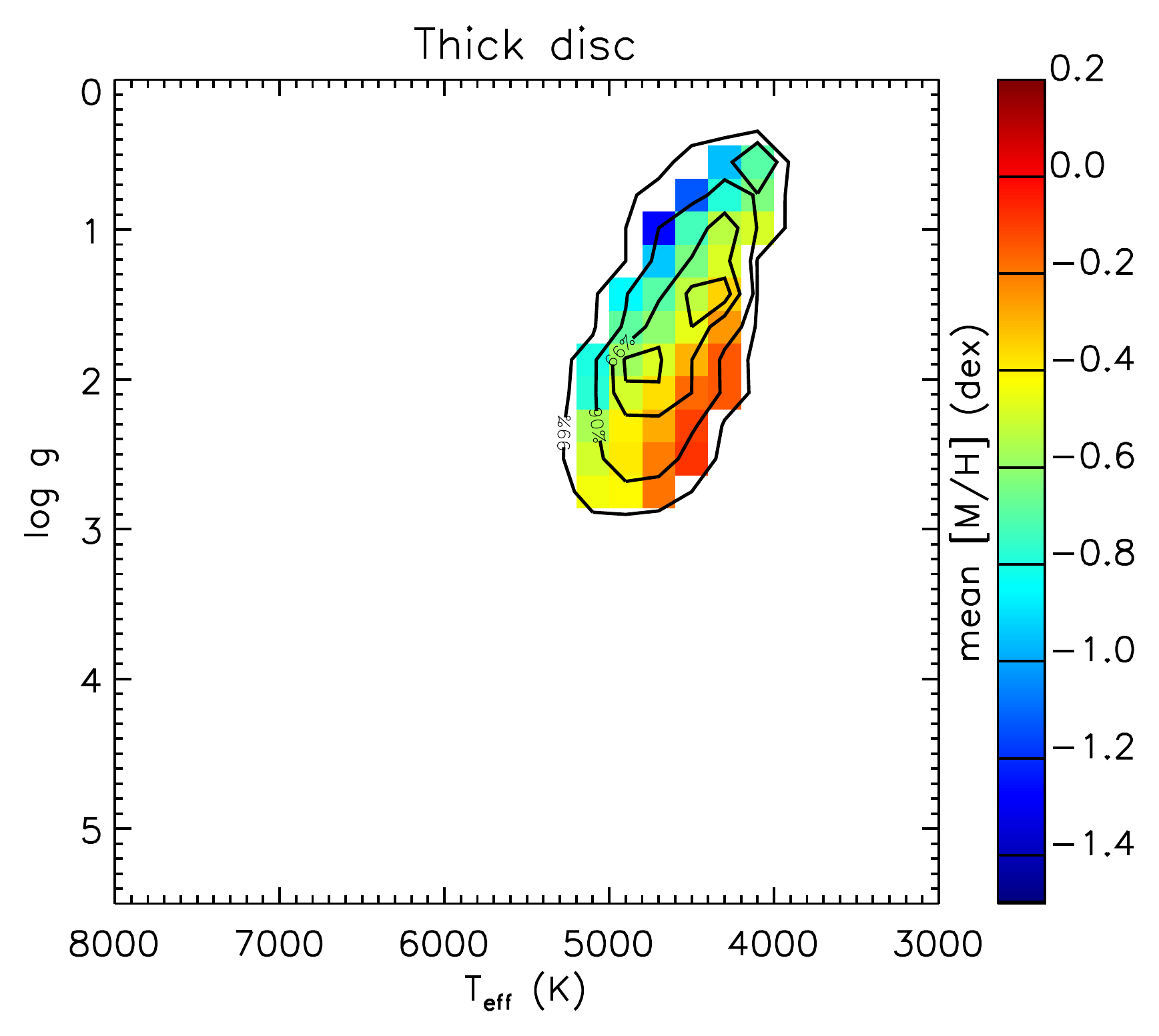}
\end{array}$
\caption{Mean metallicity across the Hertzsprung--Russell diagram of the selected stars in the Solar cylinder at distances from the Galactic plane between $1<|Z|<2$~kpc (see Sect.~\ref{sect:thick_disc_Solar_neighbourhood}).  The iso-contour lines contain 33\%, 66\%, 90\% and 99\% of the total selected sample.  }
\label{fig:HR_thick disc}
\end{figure}

\begin{figure}
\centering
$\begin{array}{c}

\includegraphics[width=0.8\linewidth, angle=0]{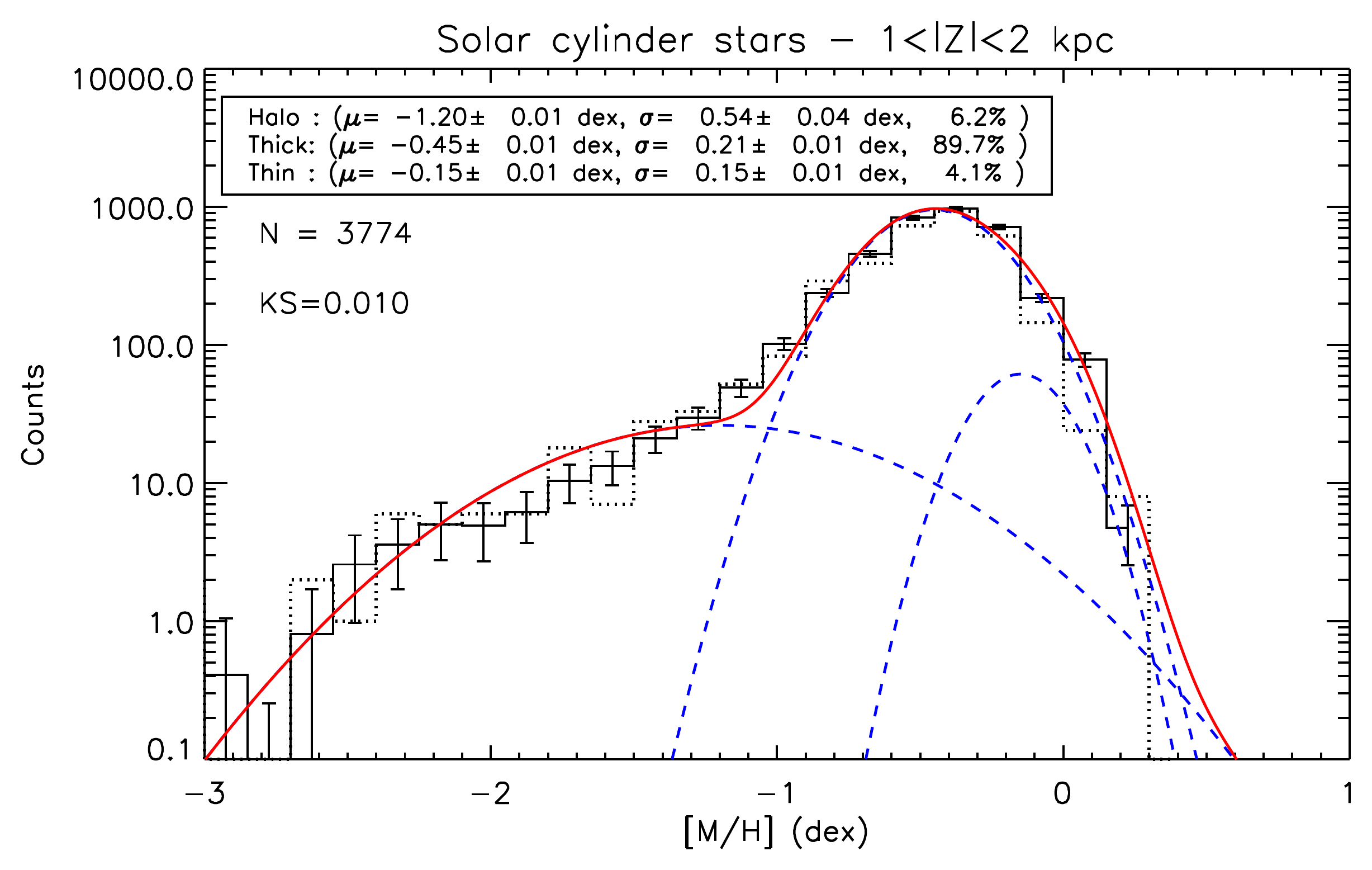}\\
\includegraphics[width=0.8\linewidth, angle=0]{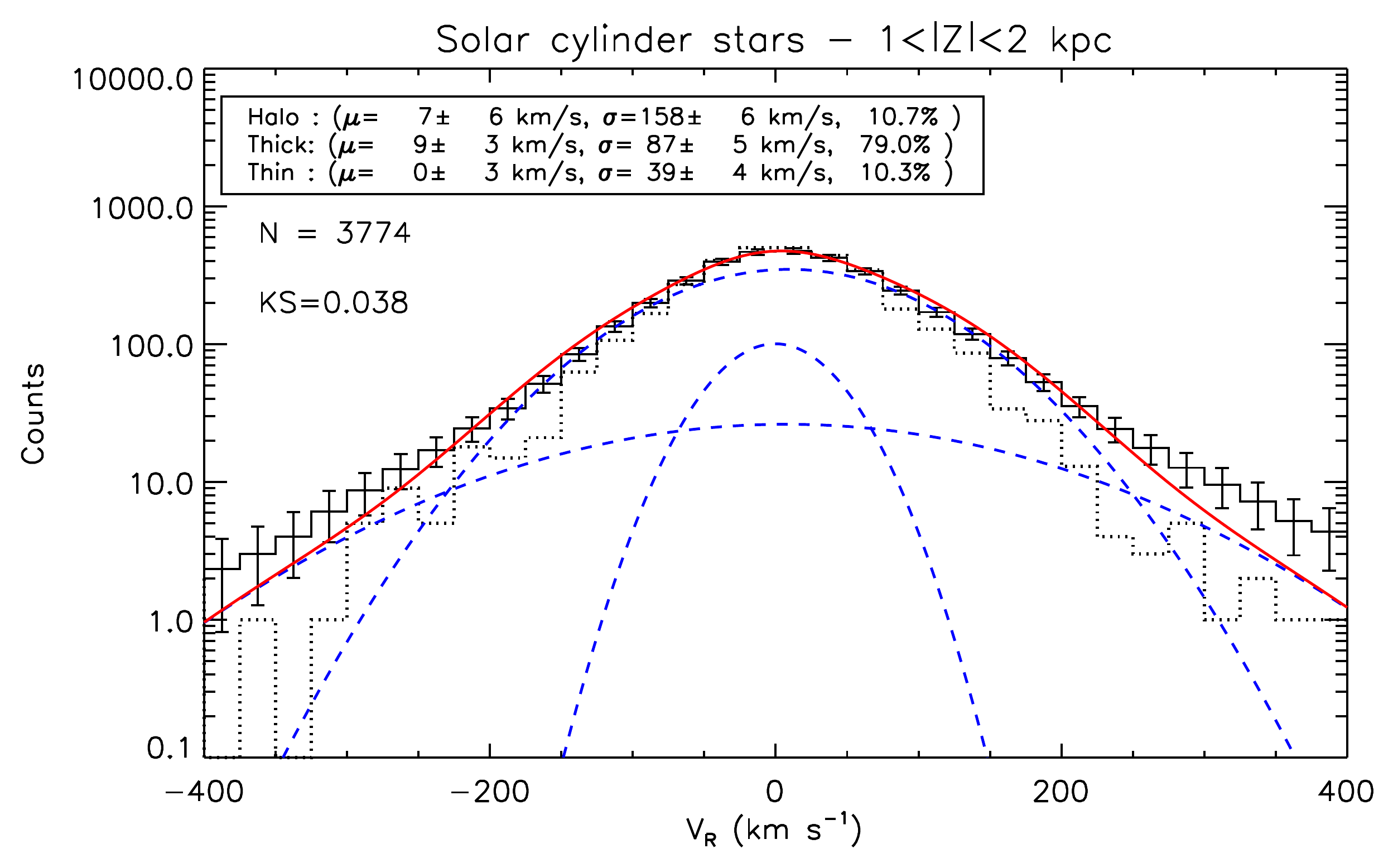}\\
\includegraphics[width=0.8\linewidth, angle=0]{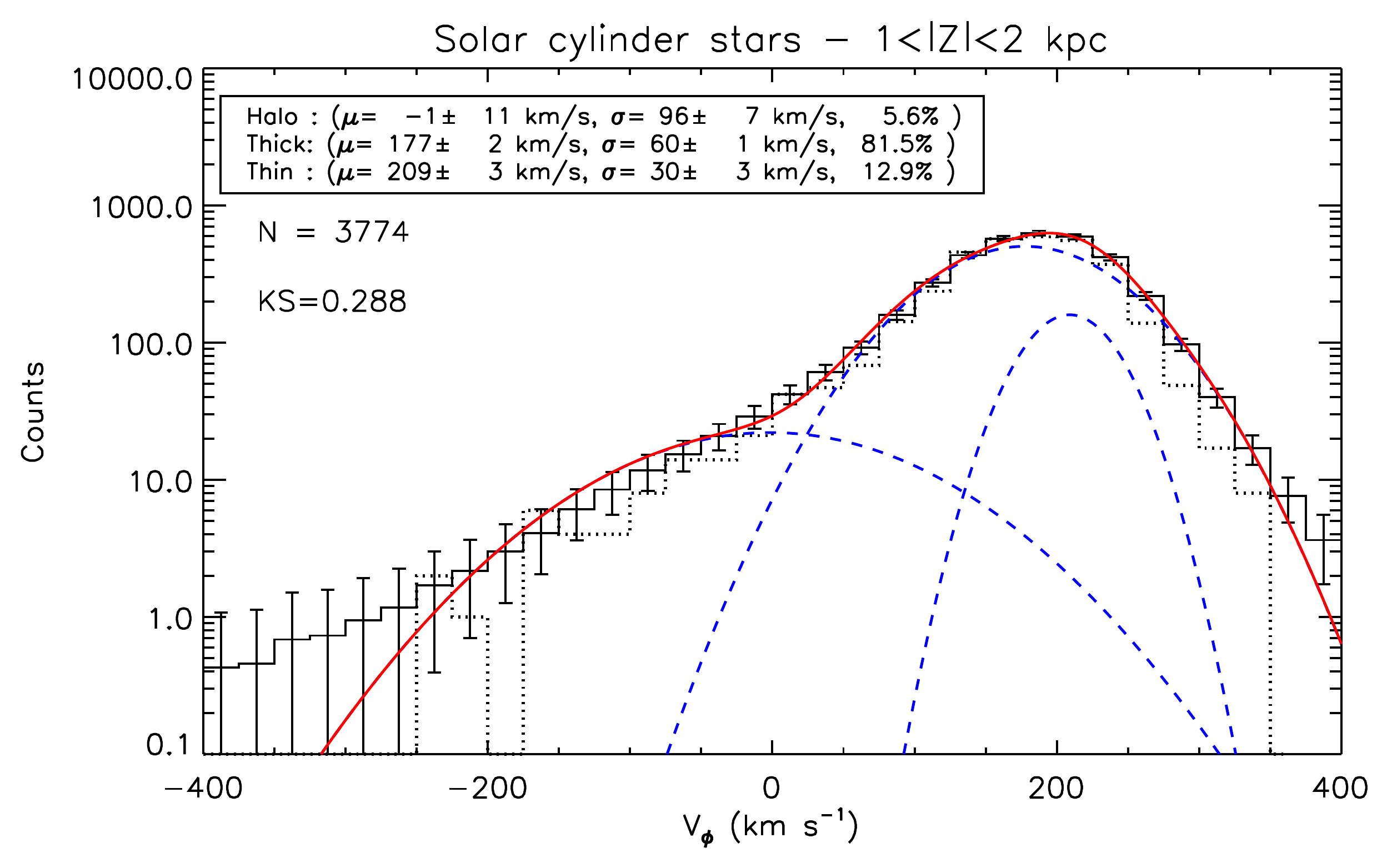}\\
\includegraphics[width=0.8\linewidth, angle=0]{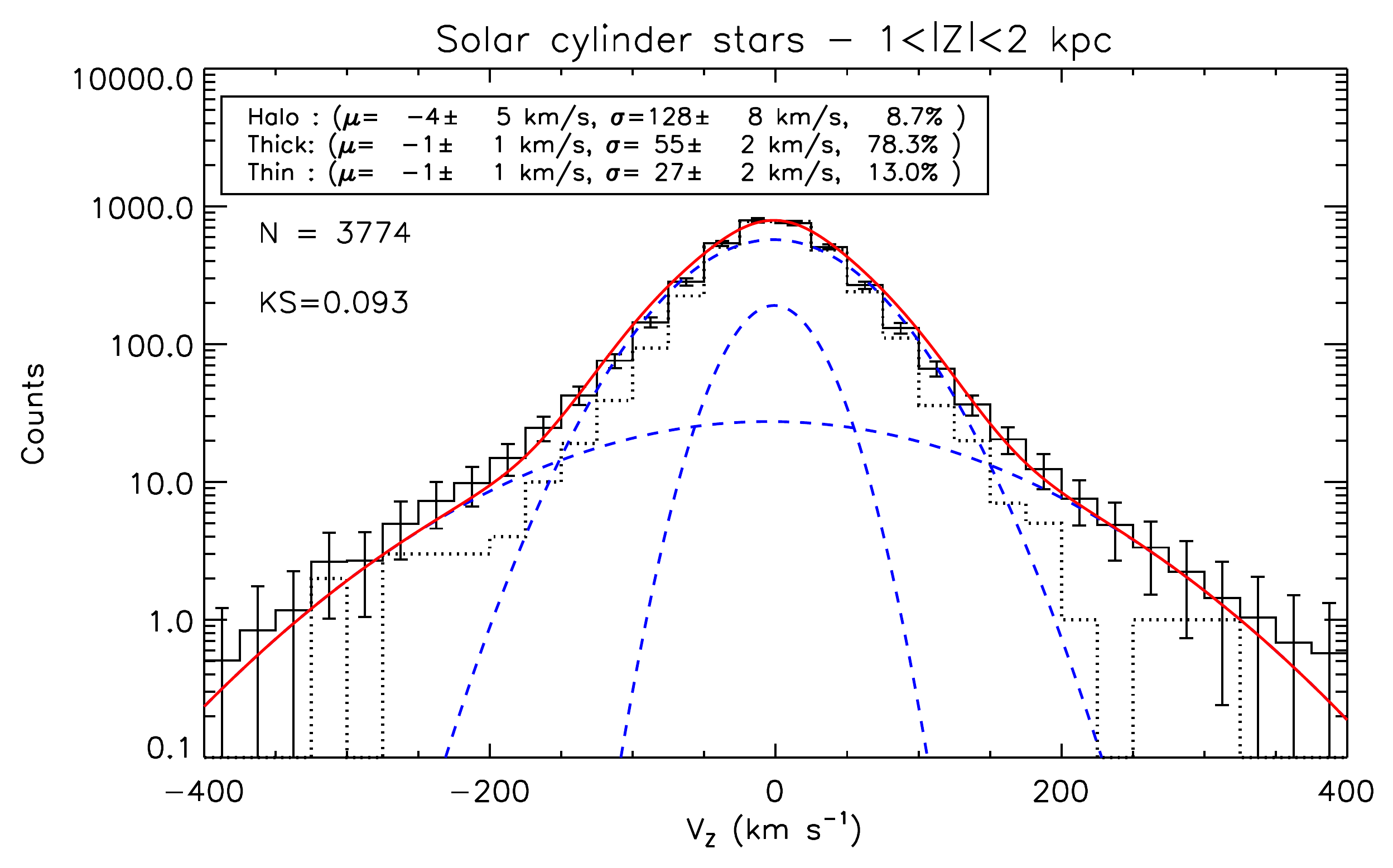}

\end{array}$
\caption{Metallicity ([M/H]), orbital Galactocentric radial velocity ($V_R$), orbital azimuthal velocity ($V_\phi$), and orbital vertical velocity ($V_Z$) distributions in the Solar cylinder, for all the stars between 1~kpc and 2~kpc away from the Galactic plane. The colour-coding is the same as in Fig.~\ref{fig:Sol_cylinder}.}
\label{fig:Sol_cylinder_thickdisc}
\end{figure}

We continued our analysis by selecting only the stars lying at $Z-$distances between one and two kpc far from the Galactic plane, at the same radial range as the previous sample ($7.5 < R < 8.5$~kpc). According to Galactic models like the one of Besan\c{c}on \citep{Robin03}, such a selection should result in a majority of thick disc stars, with roughly 10\% of stars belonging to the thin disc and 10\% to the halo. In addition, the small range in $Z$ minimises the effect of the vertical metallicity gradients of $\partial \meta / \partial Z \approx -0.1$~dex~kpc$^{-1}$ that can possibly be present in the thick disc \citep{Ruchti10, Katz11, Kordopatis11b, Kordopatis13a}.   The HR diagram of the selected sample is illustrated on Fig.~\ref{fig:HR_thick disc} and shows that the probed targets are mainly giant stars.The resulting metallicity and velocity histograms are shown in Fig.~\ref{fig:Sol_cylinder_thickdisc}.

The mean fitted values summarised in Table~\ref{table:measured_vels_sol_cyl_thick_disc}  indeed confirm that the thick disc is the dominant population of the selected sub-sample, representing $\sim 80\%$ of the total number of stars. In addition,  for each Galactic component the mean azimuthal velocity and the metallicity values are compatible with the ones determined from the entire sample (see Table~\ref{table:measured_vels_sol_cyl} and Fig.~\ref{fig:Sol_cylinder}), 
 showing no indication of any vertical gradient on the intrinsic chemo-dynamical  properties of the thick disc. We note, however, that the velocity dispersions of the thin and thick discs are higher now by roughly 20--30~\kms. This reflects both an astrophysical behaviour
 \citep[stars at higher altitude have larger dispersions, see][]{Williams13}, 
 but is also indicative of the larger errors in the individual velocity measurements, which tend to naturally increase the intrinsic dispersions. Despite these differences, the overall agreement shows the consistency of our approach: as a first approximation, the velocity DFs, as well as the MDFs can be modelled as simple Gaussians.
Finally, the absence of any significant change in the means of distributions that is found in this study  is in agreement with \citet{Kordopatis11b,Kordopatis13a} who claimed that the measured vertical gradients in the Solar cylinder are likely due to the change in the ratio of the Galactic populations with distance from the plane.

It should be noted at this stage, that we also investigated the possibility of having a bias in the distances and hence in the velocities, due to the spectroscopic degeneracy around the calcium triplet, {\it i.e.} the wavelength range of the RAVE spectra. This degeneracy correlates the atmospheric parameters in a way that, along the giant branch, the spectrum of a star having a specific $\teff$, $\logg$ and $\meta$ can be similar to another having a simultaneously higher or lower effective temperature, surface gravity and metallicity \citep[see][]{Kordopatis11a}. The correlation between metallicity and surface gravity is of the order of 1:2, {\it i.e.} a decrease of 0.2 in $\logg$ and 0.1~dex in $\meta$ is needed in order  to find identical spectra \citep[see][]{Kordopatis13b}. 
{  It is the (stochastic) position of the noise in the spectra that will determine the way the derivation of the atmospheric parameters (and thus the distances) are going to be affected. 
Tests in the derivation of the parallax by changing the input atmospheric parameters by the above mentioned values have shown that the distances are expected to increase (decrease) by roughly 5\% when decreasing (increasing) the surface gravity. However, these changes cannot be considered as a systematic, and thus } we do not expect any major influence of this effect to our sample selection, and thus on its derived properties.

\begin{figure}
\centering
$\begin{array}{c}
\includegraphics[width=0.85\linewidth, angle=0]{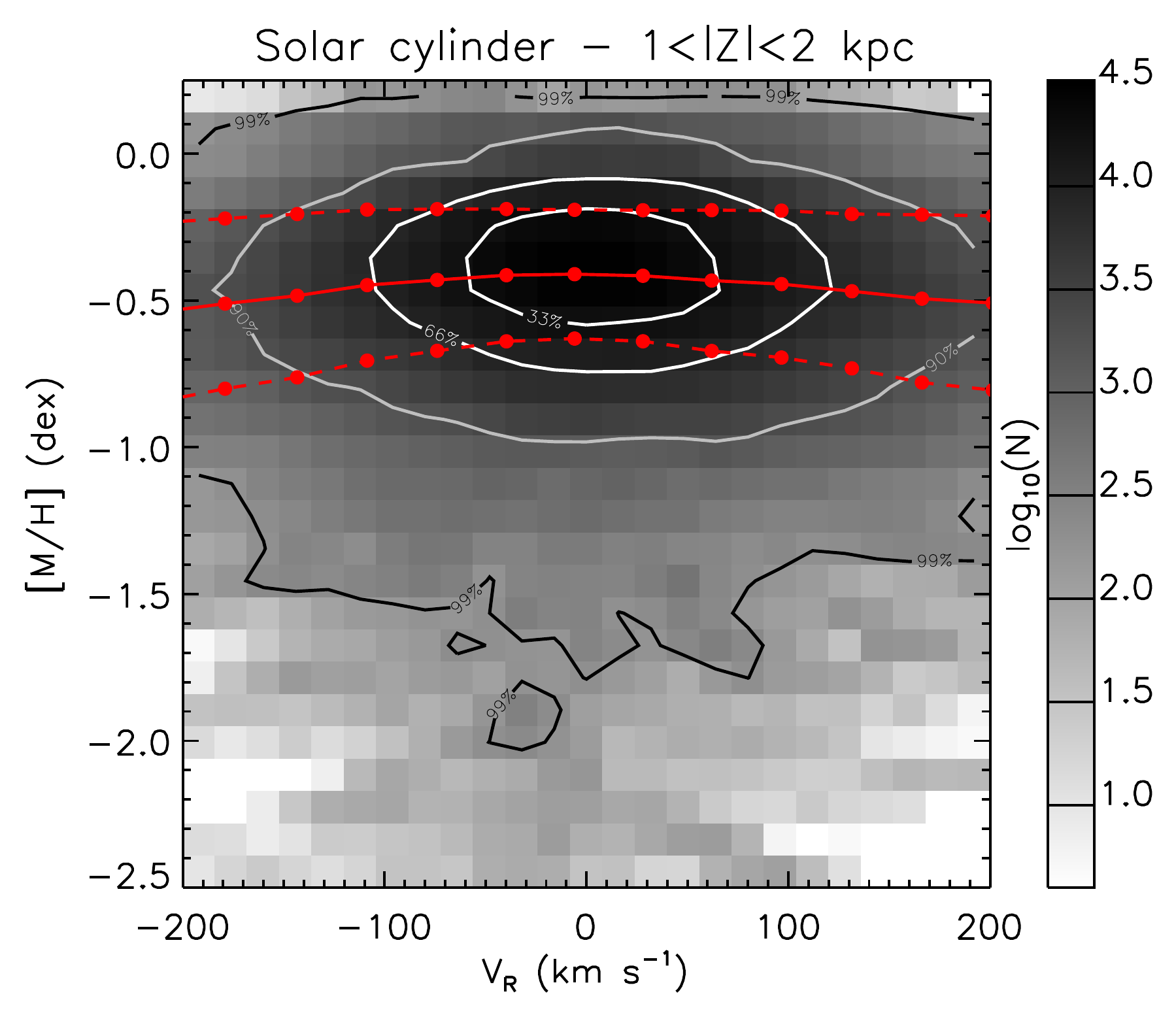} \\
\includegraphics[width=0.85\linewidth, angle=0]{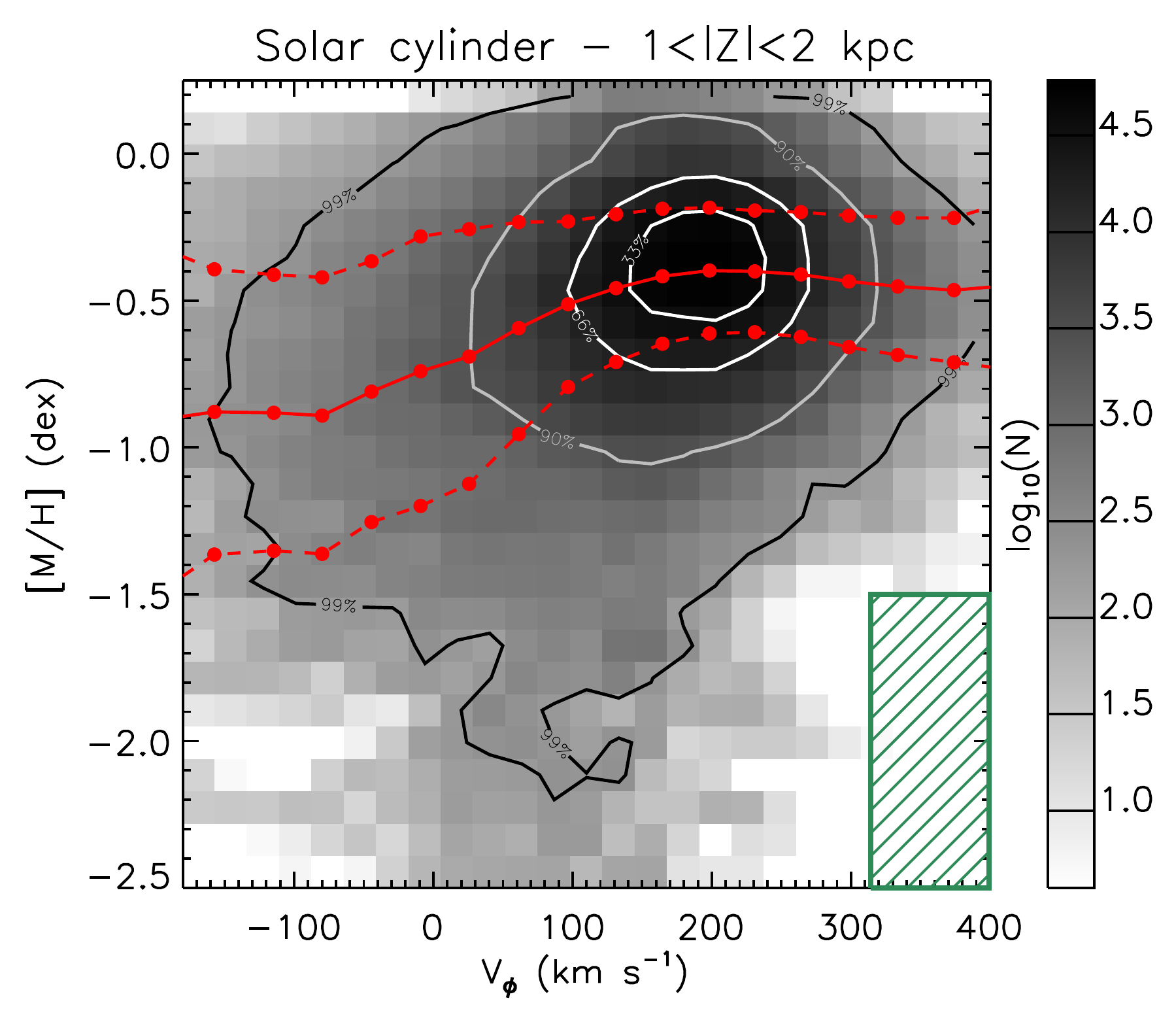} \\
\includegraphics[width=0.85\linewidth, angle=0]{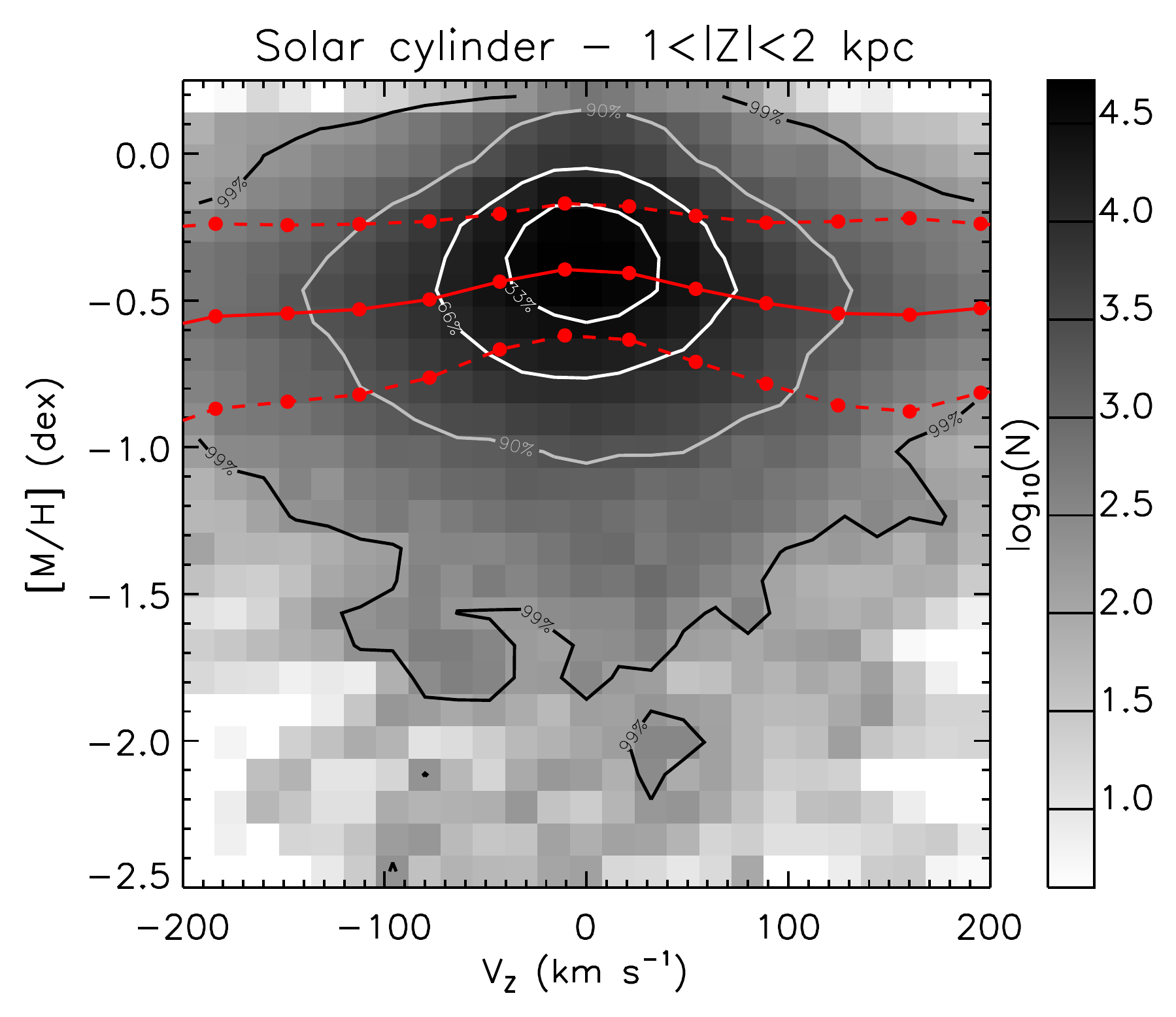} \\
\end{array}$
\caption{Correlation between the Galactic radial ($V_R$, top panel), azimuthal ($\vphi$, middle panel) and vertical ($V_Z$, bottom panel) velocities, and  the metallicity, $\meta$, for the stars in the Solar cylinder and $1<|Z|<2$~kpc. The grayscale colour coding represents the logarithmic star-count per bin of 0.1~dex $\times$ 20~\kms\ after 500 Monte-Carlo realisations. The iso-contour lines contain 33\%, 66\%, 90\%, 99\% of the considered sample. The plain red lines represent the median metallicity values at different velocity values, (constantly spaced by 30~\kms), and the dashed lines are the associated dispersions. The dashed box at low metallicity and $\vphi >$310~\kms\ represents the local escape velocity for an isotropic population, assuming $V_{esc}=544$~\kms.  }
\label{fig:Vphi_meta_correl_TD}
\end{figure}

Figure~\ref{fig:Vphi_meta_correl_TD} illustrates the correlations between the metallicity and the velocities for the stellar sample considered to be thick disc dominated ($7.5 < R < 8.5$~kpc, $1<|Z|<2$~kpc). On one hand, the radial and vertical velocity components (top and bottom panels) show no particular correlations with metallicity. On the other hand, as far as the azimuthal velocity is concerned (middle panel) we find: {\it (i)} no correlation for velocities larger than the thick disc lag ($\vphi \gtrsim 180$~\kms), {\it (ii)}  a positive slope for $\vphi$ smaller than 180~\kms, and {\it (iii)} a flat trend for the counter-rotating stars.

The interpretation of the above three regimes can be done in the following way: 
the thick disc has an intrinsic correlation between metallicity and $\vphi$ of the order of $40-50$~km~s$^{-1}$~dex$^{-1}$, which is at the origin of the slope seen at  $-1 < \meta < -0.5$~dex and $0 < \vphi < 180$~\kms. At higher velocities, the increasing proportion of the thin disc stars leads to the flat trend that is observed. Indeed,  \citet{Lee11} have shown that there is an anti-correlation between the azimuthal velocity and the metallicity of the order of $-30$~\kms ~dex$^{-1}$ (see their Fig.~7) for the thin disc stars. The combination of these two opposite correlations causes the flat trend that is seen in Fig.~\ref{fig:Vphi_meta_correl_TD}.  
Towards azimuthal velocities lower than 150~\kms, the proportion of the halo stars is increasing, which tends to lower the mean metallicity, and hence increase the slope.
Finally, the flat trend for the counter-rotating stars is due to the fact that the majority of these stars belong to the halo, which is not expected to have any correlation between its kinematics and its chemistry.

\begin{figure}
\centering
$\begin{array}{c}
\includegraphics[width=0.9\linewidth, angle=0]{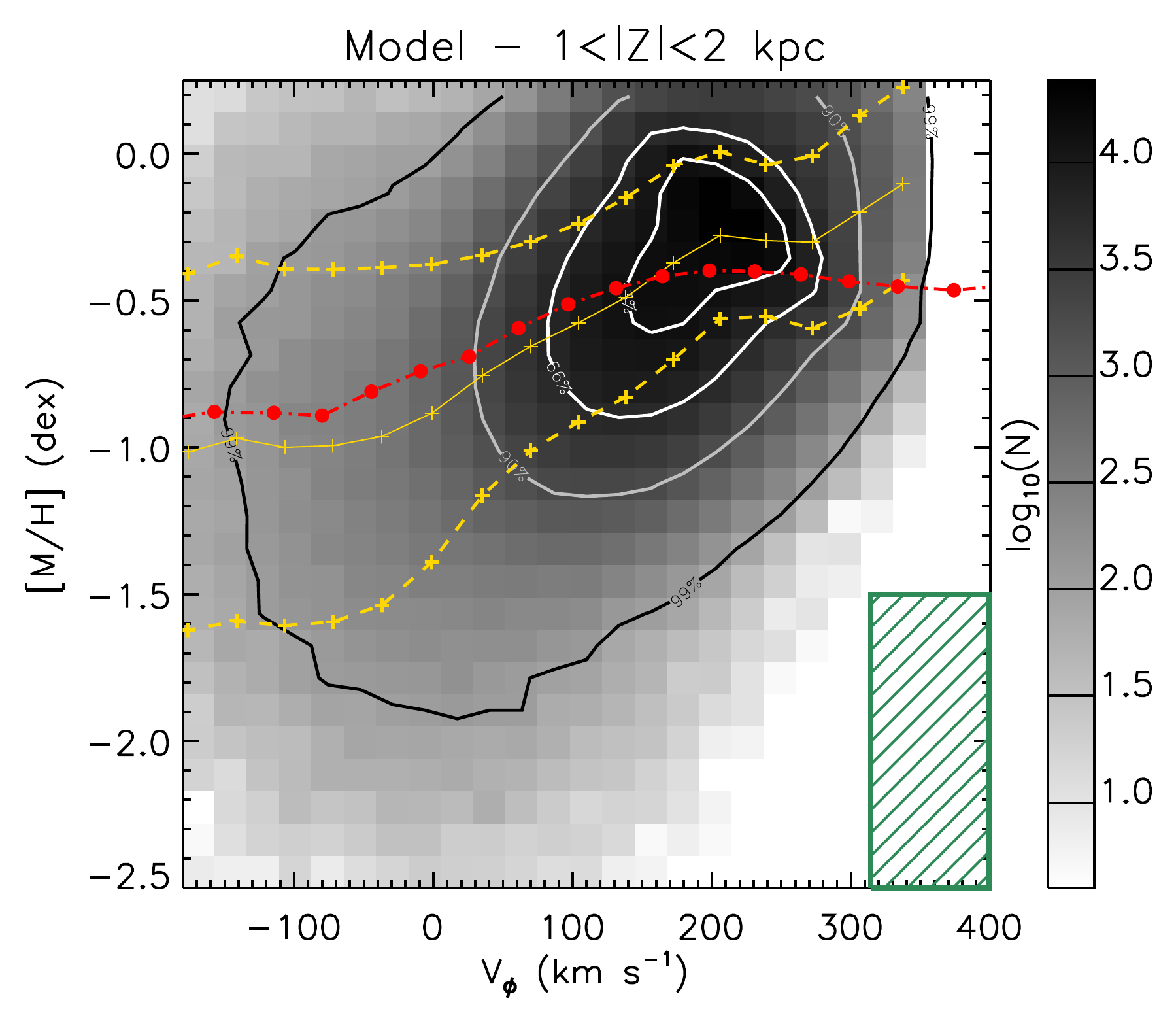} 
\end{array}$
\caption{Same as the middle panel of Fig.~\ref{fig:Vphi_meta_correl_TD} but for a our simple three component Gaussian model. The red curve is the result obtained from the observations, {\it i.e.} the one from the middle panel of Fig.~\ref{fig:Vphi_meta_correl_TD}.}
\label{fig:Vphi_meta_correl_TD_model}
\end{figure}

Figure~\ref{fig:Vphi_meta_correl_TD_model} confirms the validity of our above stated interpretation. In this plot, we have simulated the chemo-dynamical properties of $10^5$~stars. We have modelled the thin disc, the thick disc and the halo as having Gaussian  metallicity and azimuthal-velocity distribution functions. The halo has a metallicity independent of its velocity, the thin disc a correlation of $-30$~\kms ~dex$^{-1}$ and the thick disc a correlation of $40$~\kms ~dex$^{-1}$.  The mean metallicity of the halo is set at $-1$~dex to match the mean metallicity of the counter-rotating stars (see Sect.~\ref{sect:halo}), the mean metallicity  of the thick disc is $-0.5$~dex, and the one of the thin disc is $-0.10$~dex. The relative proportions of each population are the ones found in Table~\ref{table:measured_vels_sol_cyl_thick_disc}. One can see from that plot that the shape of the trend of the mean metallicities for velocities lower than $180$~\kms\ reproduce the observations well  (red curve on Fig.~\ref{fig:Vphi_meta_correl_TD_model}). For the {\it thin disc-like} velocities ($180 < \vphi \lesssim 270$~\kms), we can reproduce the flat trend, although with a slight offset in the mean metallicities.  More interestingly, we find a disagreement between the model and the observations for $\vphi\gtrsim 270$~\kms.  On the one hand, the origin of the disagreement with the observations is due to the fact that the assumed velocity distributions are Gaussian , whereas in reality they are skewed, and thus not many stars should have these velocities, unless their velocity errors are large, like it is discussed in Sect.~\ref{sect:Solar_cylinder}. 

\begin{figure*}
\centering

$\begin{array}{c}
\includegraphics[width=0.9\linewidth, angle=0]{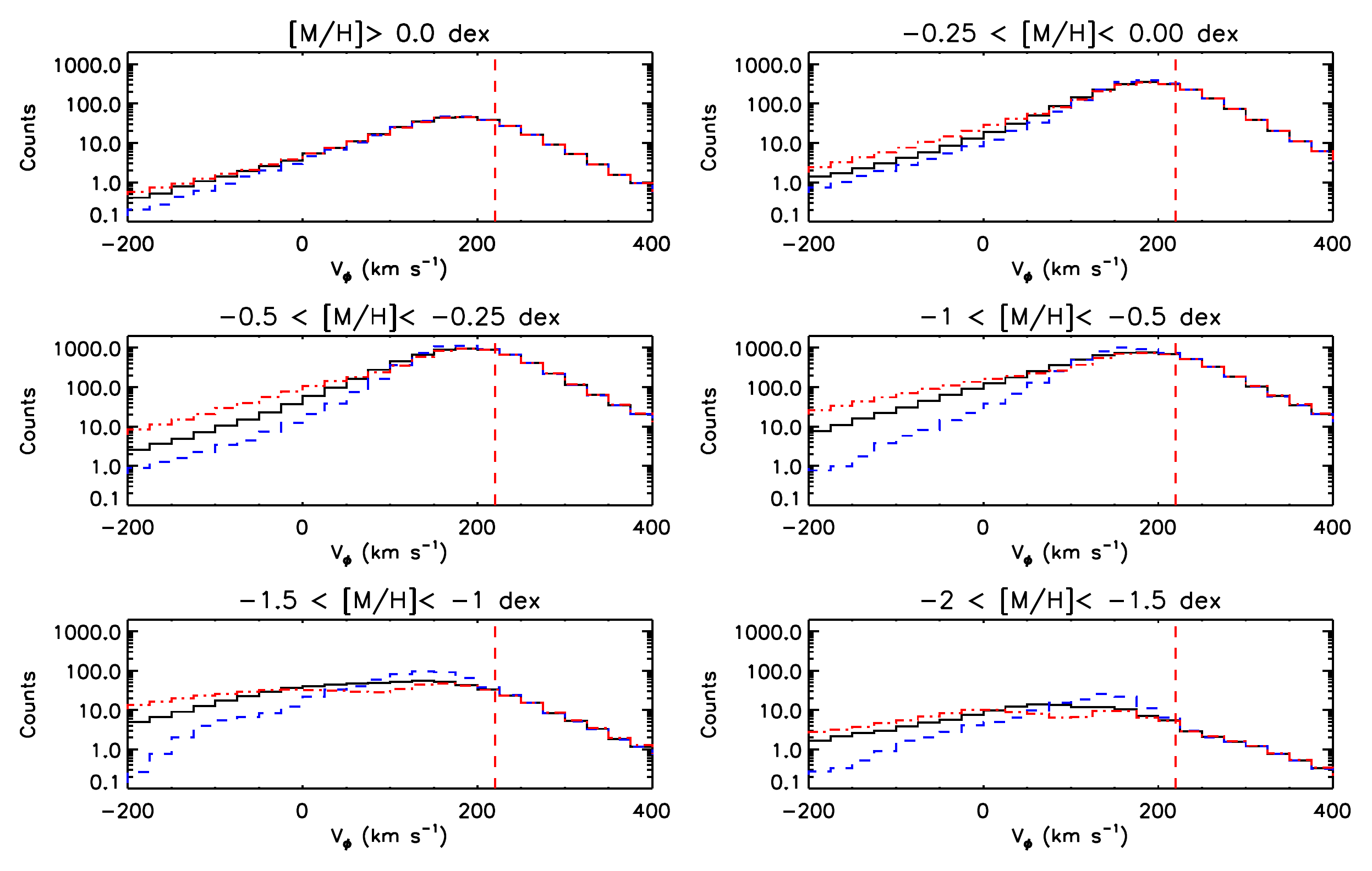}
\end{array}$

\caption{Azimuthal velocity histograms for different metallicity bins and for the stars being between 1 and 2~kpc above the Galactic plane. The counts are averaged on the total of the 500 Monte-Carlo simulations. Black histograms represent the result obtained with the actual derived distances, whereas the blue and the red ones are the results obtained with distances modified by $-50\%$ and $+50\%$, respectively for the stars with $\vphi<100$~\kms {  \citep[which is an over-estimation of the possible distance bias of 20\% for the giant stars, see][]{Binney13b}}. The vertical red dashed line is positioned at $V_{LSR}=220$~\kms.}
\label{fig:Vphi_meta_histograms_TD}
\end{figure*}

Figure~\ref{fig:Vphi_meta_correl_TD_model}  also indicates that the relatively large number of low angular momentum ($\vphi < 100$~\kms) and high metallicity ($\meta > -1$~dex) stars that is revealed at the the middle panel of Fig.~\ref{fig:Vphi_meta_correl_TD} is likely due to large errors on both the metallicity and the velocity. However, we pursued the investigation in order to learn 
 whether these stars are real, or due to uncertainties on the distances or proper motions.   
 {   The \citet{Binney13b} analysis of RAVE stellar kinematics found that the most distant stars could presumably suffer by distance biases up to 20\%.
For that reason,  we investigated possible biases much larger than the possibly estimated one, and }
we modified by $\pm 50\%$ the distances of the stars having $\vphi < 100$~\kms\
  and ran once more the Monte-Carlo simulations. 
 Figure~\ref{fig:Vphi_meta_histograms_TD} shows that the resulting star-counts are not much altered for the highest metallicities ($\meta > -0.25$~dex), whereas they change by an order of magnitude for the lower metallicities.  
 This behaviour suggests that the counter-rotating metal-rich stars are due to large uncertainties on either the distances or the proper motions, whereas the low angular momentum stars with the intermediate metallicities  should be real, as long as the distances have no significant biases (see B13).

Finally, one can also notice on Fig.~\ref{fig:Vphi_meta_correl_TD} that  the azimuthal velocity distribution for the lower metallicity stars is not centred at 0~\kms, as expected for a halo-dominated population (see Fig.~\ref{fig:Vphi_meta_correl_TD_model}). {  This feature is robust to distance biases up to 40--50\%, and} as we will see in Sect.~\ref{sect:metal_poor}, is due to the existence of a metal-weak thick disc.

\subsection{Verification of the parameters on the halo~stars}
\label{sect:halo}

We further investigate the robustness  of our methodology by testing the found parameters for the halo stars. It is recalled that the halo stars are forming a pressure supported population,  with a mean azimuthal motion close to zero and a large $\vphi$\ dispersion. In addition, they are thought to be the most metal-poor stars in the Galaxy \citep[$\mu_\meta \approx -1.2$~dex, e.g.:][]{Carollo10}.

In order to explore the properties of this population, we  select all the counter-rotating stars  in the Solar neighbourhood ($7.5<R<8.5$~kpc), since almost only halo stars are thought to have these characteristics. Indeed, as we can see from Fig.~\ref{fig:Sol_cylinder_thickdisc}, considering a canonical thick disc, having a lag of 50~\kms\ and a velocity distribution function modelled as a single Gaussian with $\sigma_\vphi = 60$~\kms\ (which, as we saw above is an over-estimation of the dispersion), then we expect the contamination of thick disc stars in our counter-rotating stellar sample to be less than 0.2\%.
In addition, in order to improve our selection, we restrict the target selection to $1<|Z|<2$~kpc above the plane. The lower boundary prevents us from including potential  thin disc stars which  have severely underestimated velocity errors, and the upper $Z$ boundary helps minimising any Malmquist bias which could affect stars at large distances (however, we note that this bias should still be existent, to some extent, since the selection should be done on the line-of-sight distance rather than the $Z-$distance).

 \begin{figure}
\centering
$\begin{array}{c}
\includegraphics[width=1.0\linewidth, angle=0]{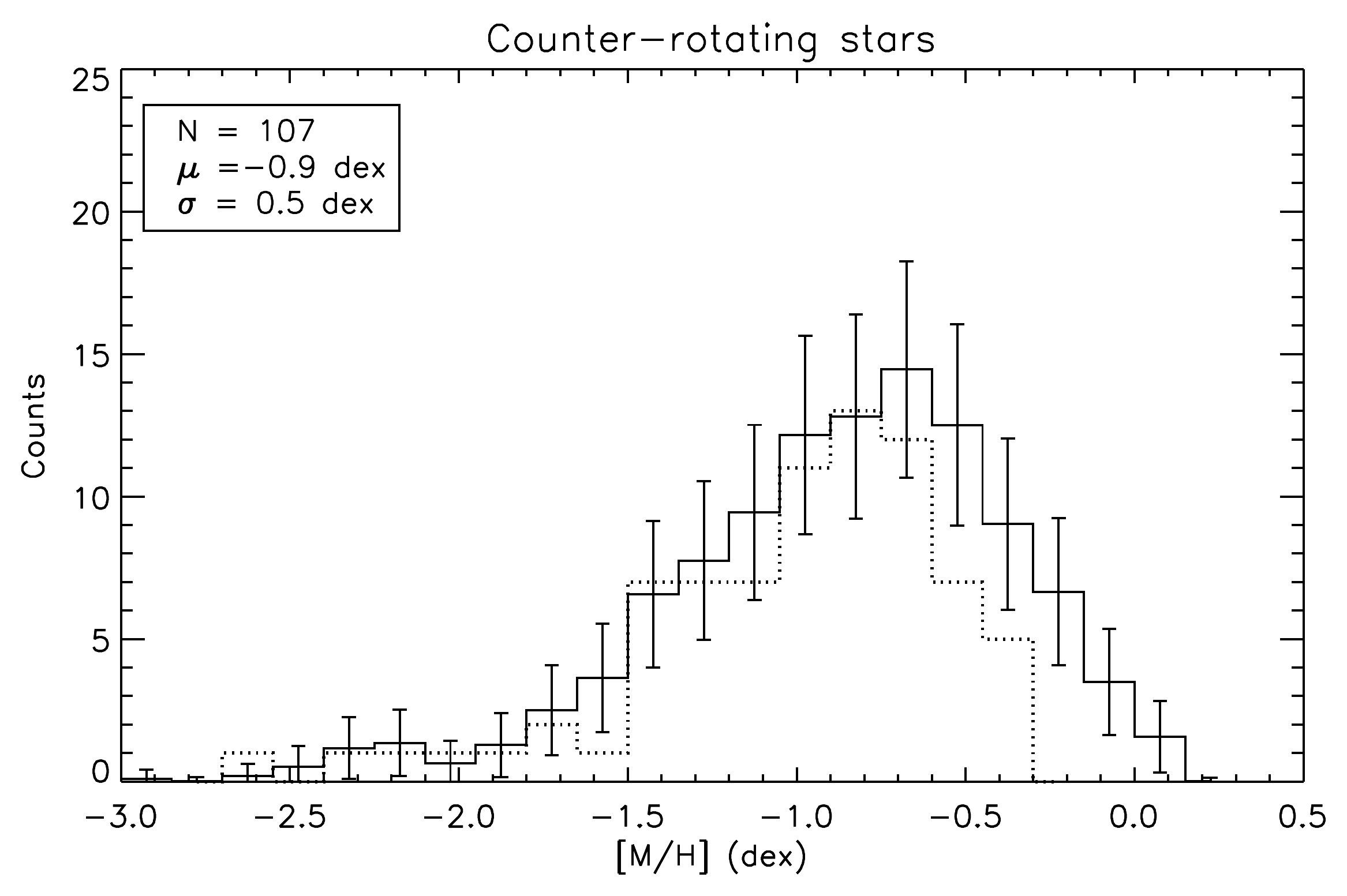}
\end{array}$
\caption{Metallicity distribution function (MDF) of the counter-rotating stars in the Solar cylinder, at $1<|Z|<2$~kpc. The dashed histogram represents the MDF of the raw measurement, whereas the plain histogram the averaged MDF obtained after 500 Monte-Carlo realisations. The total number of selected stars averaged over the Monte-Carlo simulations, their mean and standard deviation are written in the upper left corner of the figure. }
\label{fig:Counter_rotating_MDF}
\end{figure}

Statistically, over the 500 Monte-Carlo realisations there are 107 counter-rotating stars in the Solar cylinder sample (Fig.~\ref{fig:Counter_rotating_MDF}). Their mean metallicity is $\approx -0.9$~dex, which is consistent with a halo dominated population, though this value is higher compared to the mean literature value or  the one that we find when fitting the entire data-set (see Tables~\ref{table:measured_vels_sol_cyl} and \ref{table:measured_vels_sol_cyl_thick_disc}). 
The most metal-rich halo star selected by the raw measurements has $\meta = -0.3$~dex (dotted histogram of Fig.~\ref{fig:Counter_rotating_MDF}), however this metal-rich boundary value is higher when considering the Monte-Carlo realisations, due to the (Gaussian) scatter caused by the uncertainties in the metallicity measurements {  and to the fact that the raw metallicities of RAVE DR4 are slightly discretised (see K13)}. This result is consistent with the one obtained in the end of the Sect.~\ref{sect:thick_disc_Solar_neighbourhood}, where it has been shown that the observed low angular momentum metal-rich stars are due to the uncertainties on the distances and/or the proper motions.
More interestingly, by considering that the azimuthal velocity distribution function of the halo stars is even and centred on 0~\kms, then this implies that our sample has potentially 214 halo stars, representing $\sim 6\%$ of the sample in the Solar cylinder. Once again, this number is fully consistent with the results of  Table~\ref{table:measured_vels_sol_cyl_thick_disc}.

\section{Kinematic properties of the Metal-poor tail of RAVE}
\label{sect:metal_poor}

We now select the metal-poor stars having $\meta < -1.5$~dex, independently of  their  radial position, but being located between 1~kpc and 2~kpc from the Galactic plane. The considered cuts in metallicity and positions ensure us that the thin disc is in practice non-existent in our sample, so there is no need to model it.

\begin{figure}
\centering
$\begin{array}{c}
\includegraphics[width=0.9\linewidth, angle=0]{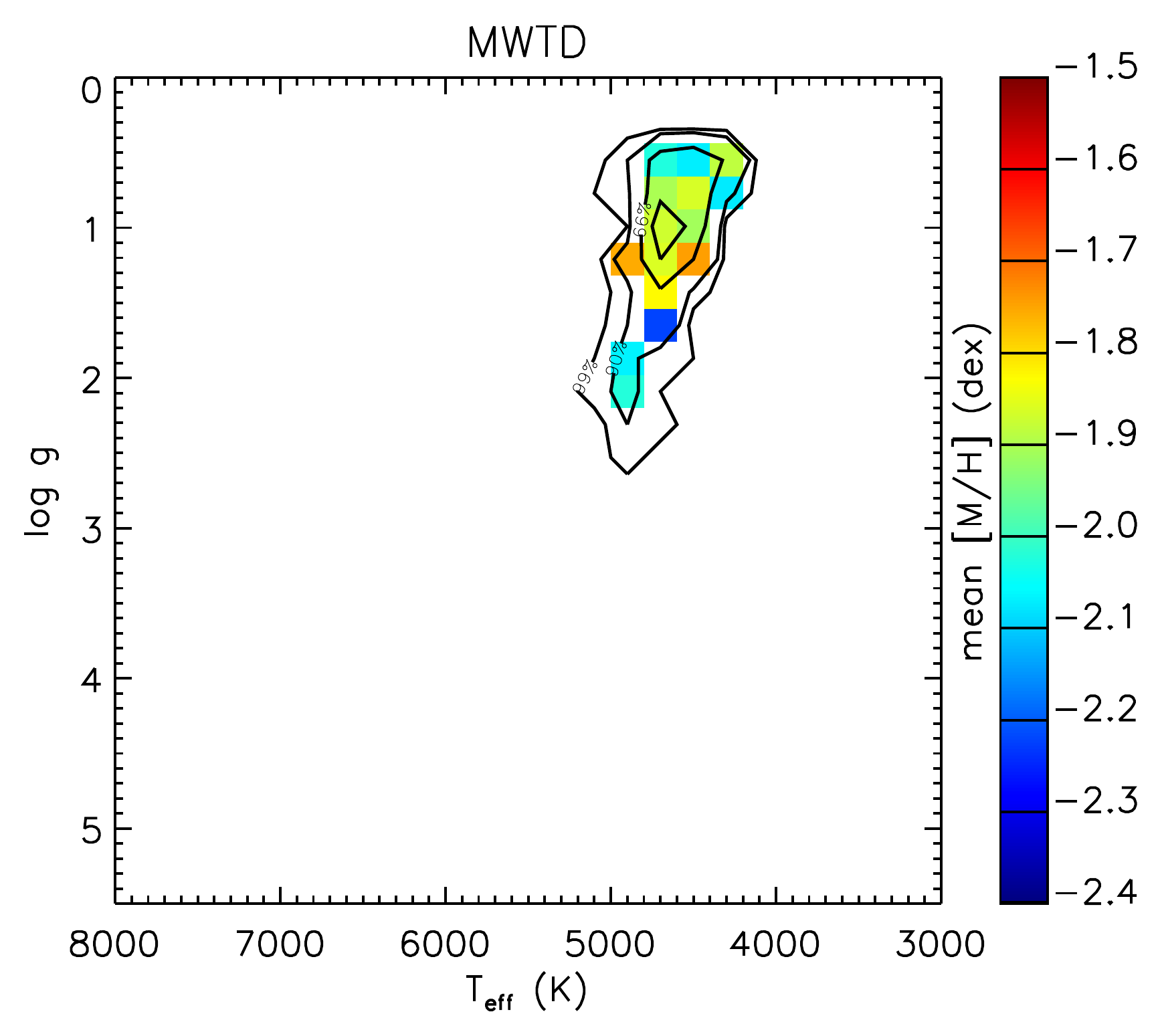}
\end{array}$
\caption{Mean metallicity across the Hertzsprung--Russell diagram of the selected stars at distances above the Galactic plane between $1<|Z|<2$~kpc  and $\meta<-1.5$~dex (see Sect.~\ref{sect:metal_poor}).  The iso-contour lines contain 33\%, 66\%, 90\% and 99\% of the total selected sample.  }
\label{fig:HR_mwtd}
\end{figure}

\begin{figure}
\centering
$\begin{array}{c}
\includegraphics[width=1.0\linewidth, angle=0]{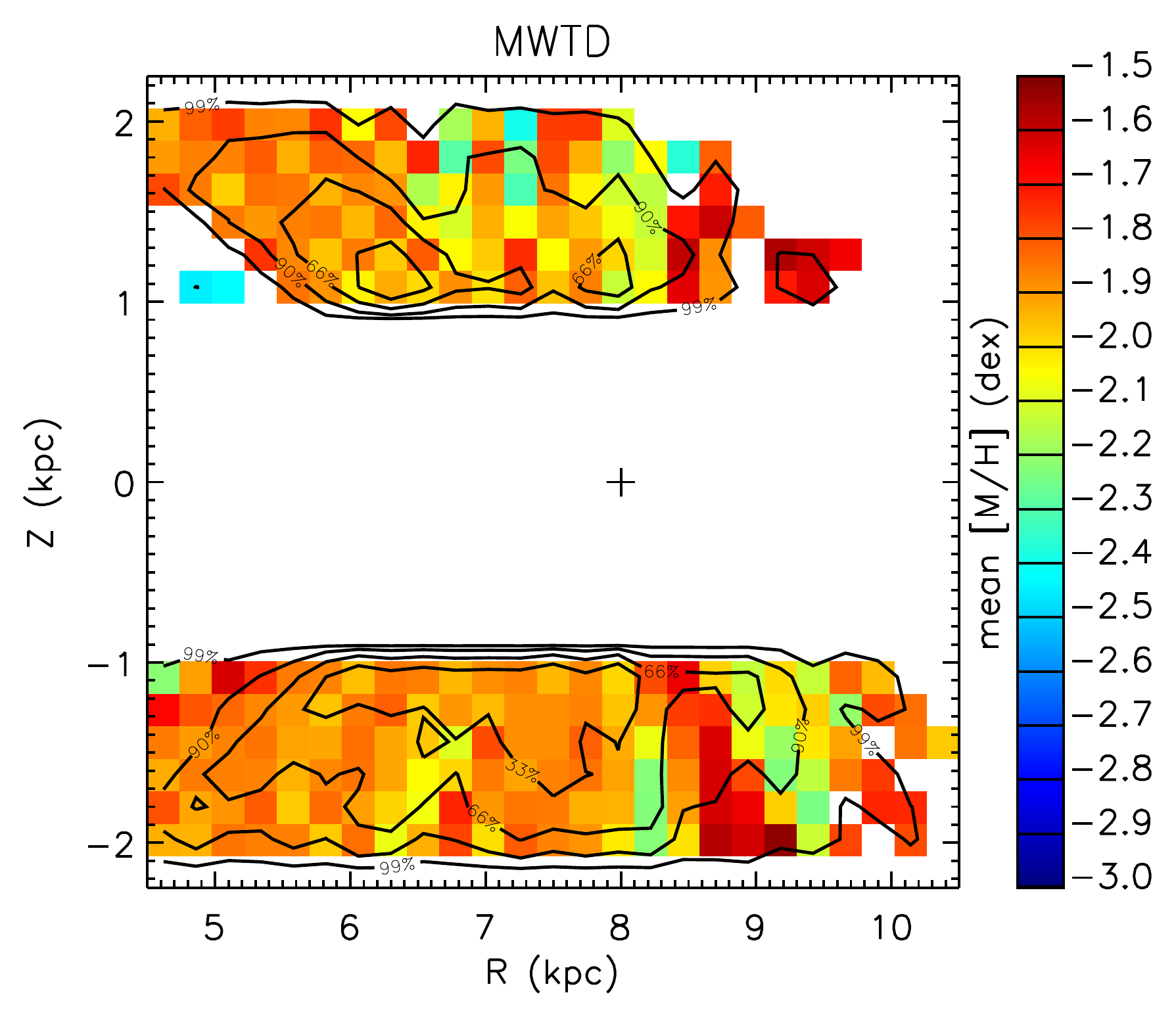}
\end{array}$
\caption{Mean positions and metallicities in the $R-Z$ plane, averaged over the Monte-Carlo realisations, for the stars  with $\meta < -1.5$~dex being between 1~kpc and 2~kpc from the Galactic plane. The position of the Sun is indicated as a $"+"$ symbol, at ($R_\odot,Z_\odot)=(8,0)$~kpc. The iso-contour lines contain 33\%, 66\%, 90\% and 99\% of the total sample. The colour-coding represents the mean metallicity inside each 0.25~kpc bin. No particular radial metallicity trends are observed in the selected sample. }
\label{fig:MP_positions}
\end{figure}

\begin{figure}
\centering
$\begin{array}{c}
\includegraphics[width=1.0\linewidth, angle=0]{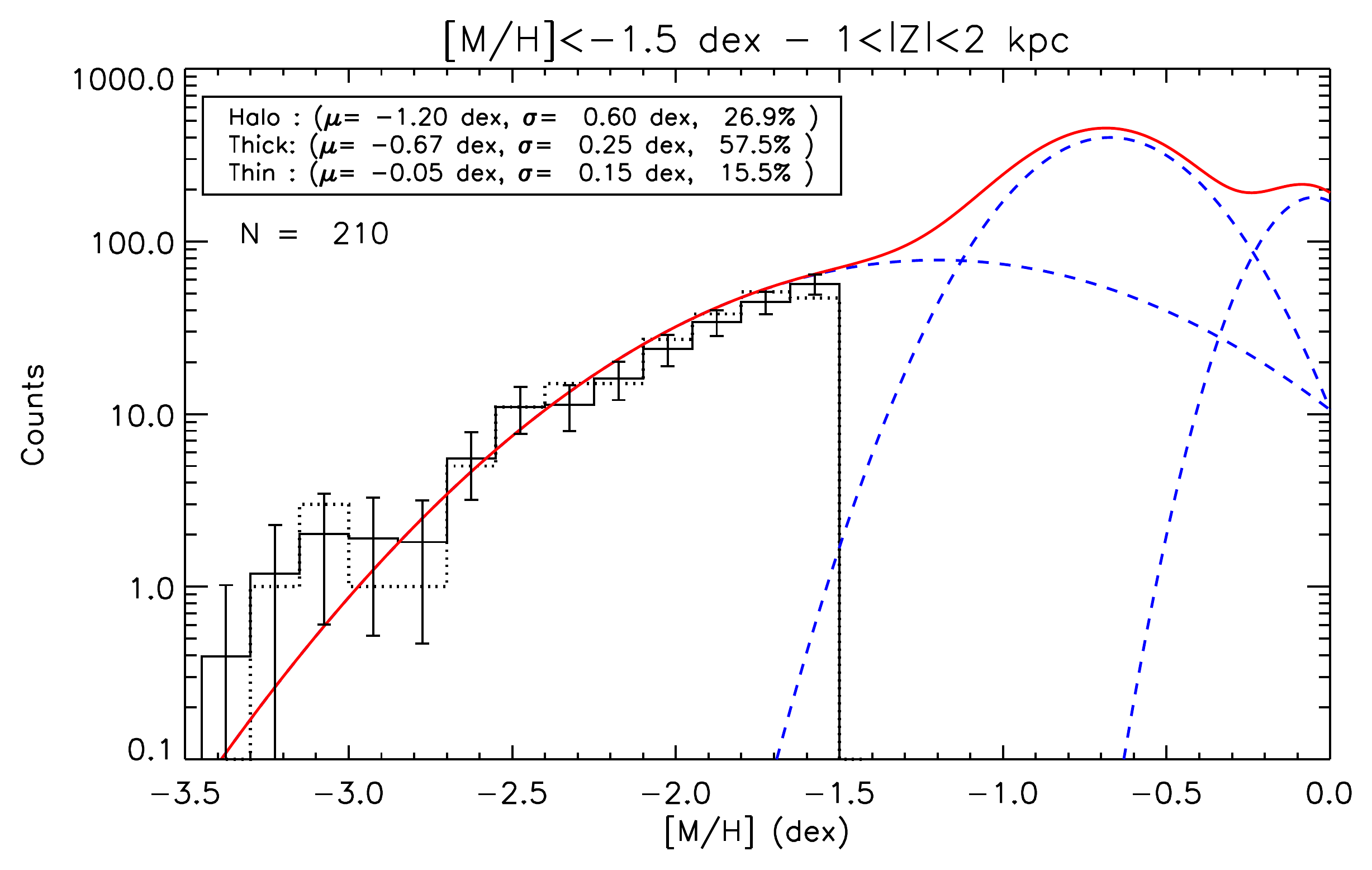}
\end{array}$
\caption{Metallicity distribution for all the stars with $\meta < -1.5$~dex and $1<|Z|<2$~kpc. The data can well be fitted with a simple Gaussian distribution having the characteristics of the stellar halo. The Gaussians corresponding to the thin and thick disc are only indicative, and have not been used to fit the metallicity distribution.}
\label{fig:MP_MDF}

\end{figure}

Statistically, less than 250 stars fulfil our selection criteria. They are giants (see Fig.~\ref{fig:HR_mwtd}), mostly spread in the inner Galaxy, and in majority towards the southern Galactic cap (see Fig.~\ref{fig:MP_positions}).
The distribution function of these metal-poor stars decreases exponentially  (Fig.~\ref{fig:MP_MDF}), as it is expected for a single Gaussian distribution associated with the stellar halo, centred at $\mu_\meta \sim-1.2$~dex with a standard-deviation of $0.6$~dex. In addition, it is worth mentioning that, as one can see from Fig.~\ref{fig:MP_MDF}, the MDF of the thick disc, modelled as a simple Gaussian centred at [M/H]$\sim -0.5$~dex and $\sigma_{[M/H]}\sim 0.25$~dex, is not expected to reach such low metallicities (less than $0.004$\% of the thick disc stars\footnote{The proportion of counter-rotating thick disc stars rises up to 0.14\% if we consider a Gaussian distribution with $(\mu_{\rm [M/H]}, \sigma_{\rm [M/H]})=(-0.6,0.3)$~dex.}).

\subsection{Azimuthal velocity distribution function: A metal-weak thick disc?}
\label{sec:MWTD_vphi}

Based on the metallicity cuts that have been applied for the target selection (Fig.~\ref{fig:MP_positions} and Fig.~\ref{fig:MP_MDF}), if the distribution functions of the thick disc were Gaussians, than the properties of the azimuthal velocities of the selected stars are expected to be the ones of the halo, {\it i.e.} $\vphi \sim 0$~\kms. Thus, any identification of stars having disc kinematics would probe the deviations from Gaussianity of both the MDF and the velocity DFs of the thick disc that we aim to highlight in this study.
  
  \begin{figure}
\centering

$\begin{array}{c}
\includegraphics[width=0.9\linewidth, angle=0]{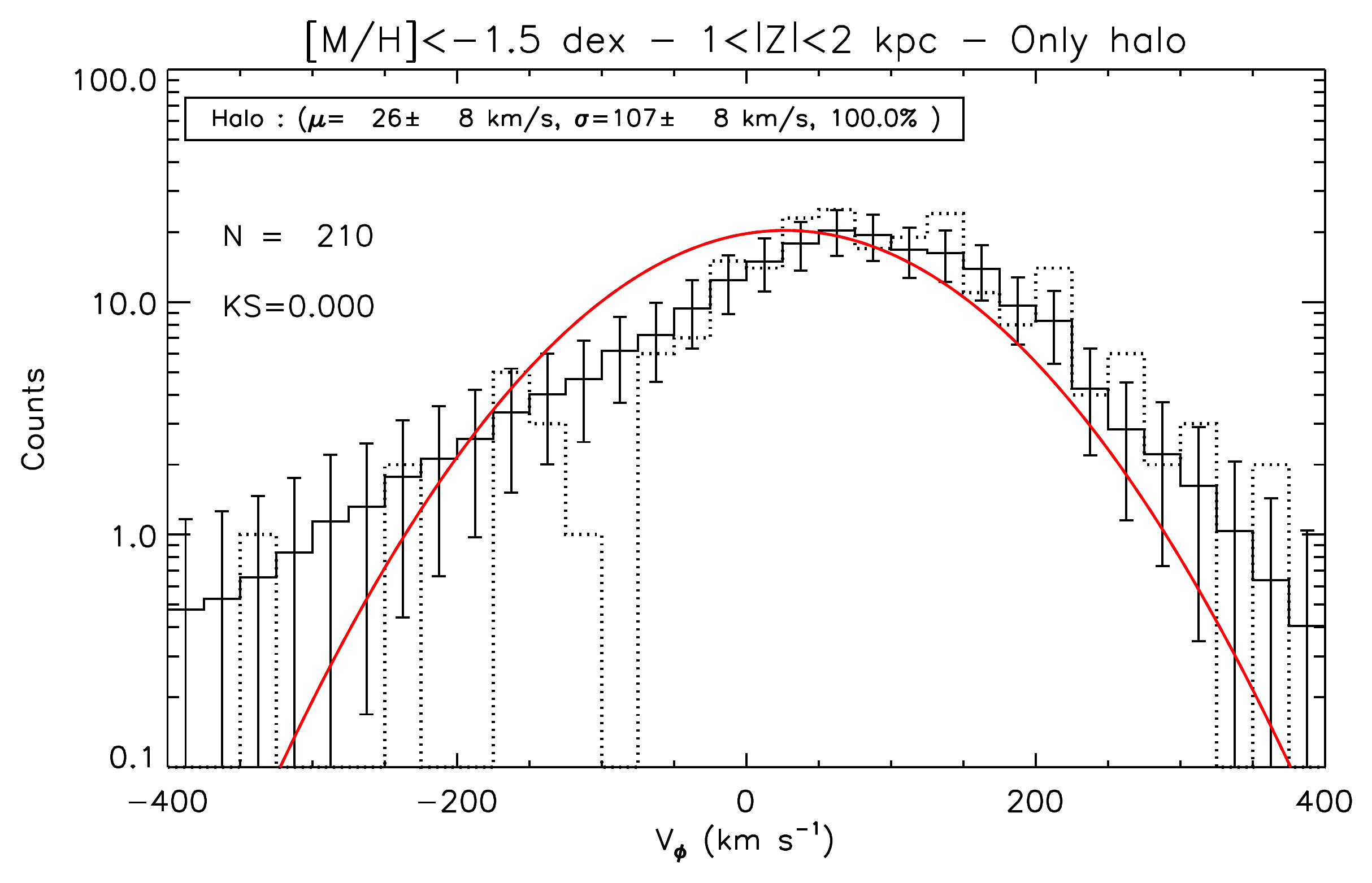}\\
\includegraphics[width=0.9\linewidth, angle=0]{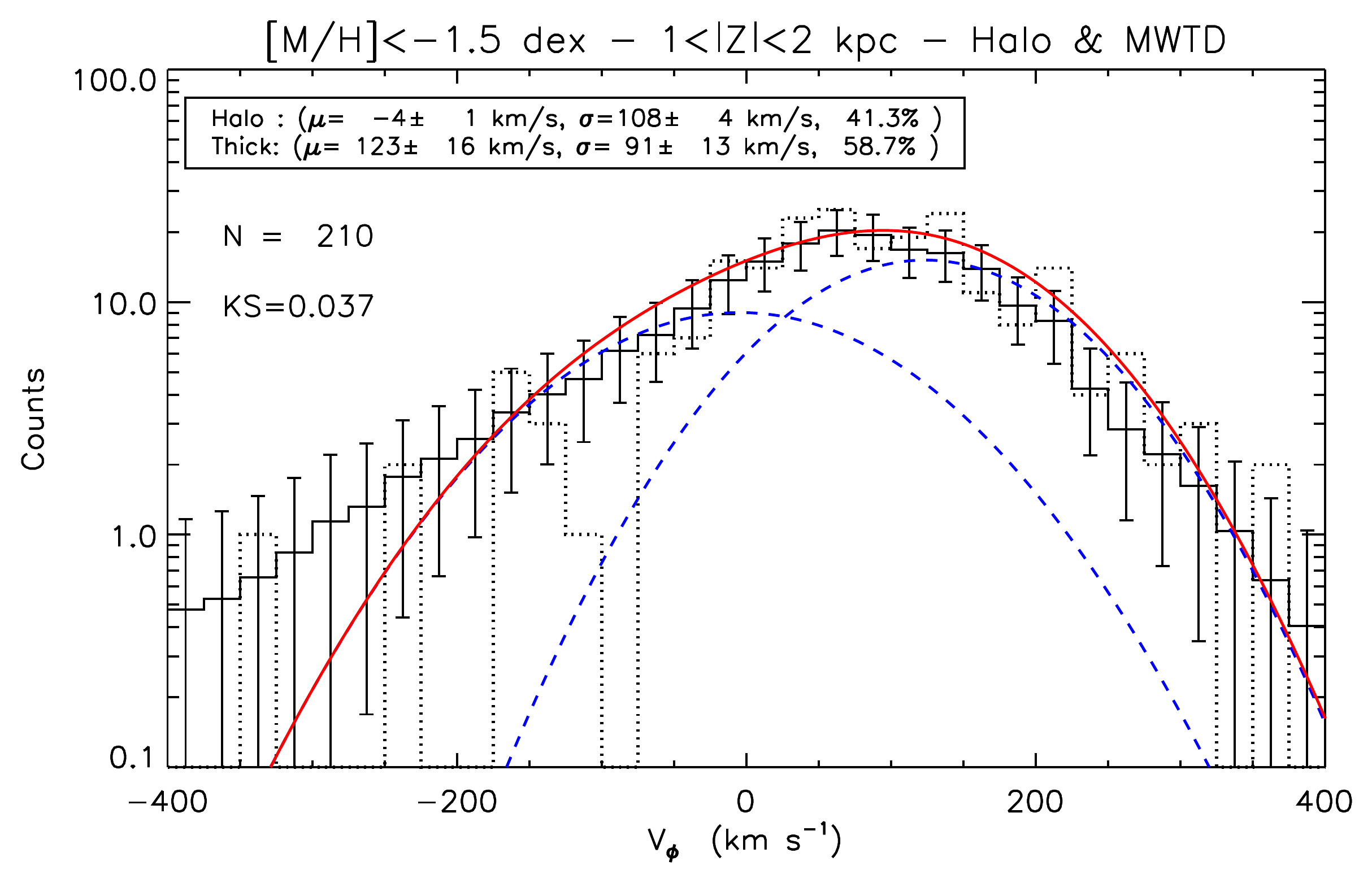}\\
\end{array}$

\caption{Azimuthal velocity histogram for all the stars with $\meta < -1.5$~dex and $1<|Z|<2$~kpc. The top plot is the result where only one Gaussian is fitted to the data, corresponding to the typical halo population. In the lower plot, two Gaussians are fitted: one corresponding to the halo, and one having intermediate values between the halo and the canonical thick disc. In order to best fit the data, a second Gaussian is required, centred at $\vphi \sim 125$~\kms.  }
\label{fig:MP_Vphi}
\end{figure}

As Fig.~\ref{fig:MP_Vphi} shows, when trying to fit  the data using our Maximum-Likelihood procedure with only one Gaussian having the halo characteristics, then the KS test rejects the null hypothesis that the model and the observations are drawn from the same underlying population. Two Gaussians are needed to find an acceptable agreement with the observations, one centred at the mean azimuthal velocity of the halo, and another, comprising roughly 60\% of the stars, centred at $\vphi \sim  120$~\kms, an azimuthal velocity close to the canonical value of the thick disc.  We highlight the fact that the histogram below $\vphi \sim -250$~\kms\ is not fitted by the model, since the counts for these bins are statistically consistent with noise. 
The velocity lag for the {\it ``extra''} component, of $V_{lag} \sim 100$~\kms~, is close to the lag measured for the metal-weak thick disc (MWTD, hereafter) found by \cite{Carollo10} using data from the SDSS, however their $\vphi-$dispersion ($\sigma_\vphi = 40$~\kms) is {  significantly lower than the one found in our analysis ($\sigma_\vphi \approx$ \svphideMWTD~\kms, or $61$~\kms\ once corrected by the error measurements, see Sect.~\ref{sect:scale_lenghts}).}

 {  
 We have investigated the distance errors and  proper motions errors  distributions for these stars to see if there were any systematics that could create such an artificial signature, though without noticing any particular suspicious pattern. 
In addition, we note that the tests that have been done on the distance estimations based on Hipparcos stars \citep{Binney13} and RAVE kinematics \citep{Binney13b} do not indicate any distance bias exceeding 20\%.   
Since Fig.~\ref{fig:Vphi_meta_histograms_TD} (bottom plots) suggests that the distances must be changed by at least $+50\%$ in order to make this high angular momentum population disappear, we hence conclude that the latter is most likely true and not a spurious result of our data-set.
 }

Finally, following \cite{Binney_Merrifield} (Chap.~11.3.2), we assess the value of the vertical velocity dispersion in the inner parts of the Galaxy, where the majority of the metal-poor targets are located. We suppose that the velocity dispersion declines exponentially with radius with a $e-$folding length of $2R_d$, where $R_d$ is the scale-length of the disc that dominates the potential\footnote{This relation applies if we consider that $\sigma_{V_Z} \propto \sqrt{\Sigma}$, with $\Sigma$ the surface density varying as $e^{-R/R_d}$ \citep{vanderKruit81}.}, thus in our case the scale-length of the thin disc 
{  \citep[$\sim 2.6$~kpc, e.g.: ][]{Juric08}. 
In addition, we consider that at $R=R_\odot$ we have $(\sigma_{V_Z})_{thick}=$\svzdeTD~\kms\ (as determined in Table~\ref{table:measured_vels_sol_cyl_thick_disc}). Given these inputs,  we can infer that at $R\sim 6$~kpc the vertical velocity dispersion of the stars of the thick disc should be  $\sigma_{V_Z} \approx 80\pm3$~\kms.}
 Since the ratios between the random motions of the different velocity components are not expected to change considerably with radius \citep[][Chap. 11.3.2]{Binney_Merrifield}, we infer that at $R=6$~kpc, $\sigma_{V_Z}\approx \sigma_\vphi$.
{  
The derived values coming from the observations are fully compatible within the errors with the estimations inferred from our simple analysis (Table~\ref{table:measured_vels_MW_thick_disc}, Fig.~\ref{fig:MP_Vphi} and top plots of Fig.~\ref{fig:MP_Vz}), implying that the high-velocity metal-poor component we observe might indeed be kinematically associated with the MWTD.}

\begin{table*}
{ 
\caption{Means, dispersions and normalisations of the measured distributions functions for the metal-poor stars at $1<|Z|<2$~kpc}}
\centering

\begin{tabular}{c c c c c c c c c c }
\hline\hline
Galactic         & $V_R$ & $\vphi$ & $V_Z$ & $\sigma_{V_R}$& $\sigma_{V_\phi}$  & $\sigma_{V_Z}$ &  $N_{V_R}$ & $N_{\vphi}$ & $N_{V_Z}$ \\
component        & \kms         & \kms             & \kms  & \kms         & \kms             & \kms & & & \\
\hline
Thick disc &\vrdeMWTD & \vphideMWTD & \vzdeMWTD & \svrdeMWTD & \svphideMWTD & \svzdeMWTD & \nvrdeMWTD & \nvphideMWTD &\nvzdeMWTD\\
Halo       &\vrhaMWTD & \vphihaMWTD & \vzhaMWTD & \svrhaMWTD & \svphihaMWTD & \svzhaMWTD  & \nvrhaMWTD & \nvphihaMWTD &\nvzhaMWTD   \\
\hline
\end{tabular}
\label{table:measured_vels_MW_thick_disc}
\end{table*}

\subsection{The case of the vertical and radial orbital velocity components}

The previous section indicated that the azimuthal velocity distribution function of the metal-poor stars of RAVE is poorly described with only one Gaussian, associated with the typical velocities of the stellar halo. Here, we investigate that argument for the two other velocity components: the radial one, $V_R$, and the vertical one, $V_Z$. Both the thick disc and the halo have distributions centred at $V_{R,Z}=0$~\kms, so the only factor distinguishing the two distributions are their dispersions. Typical values, at the Solar neighbourhood, that are found in the literature are $(\sigma_{V_R},\sigma_{V_Z})_{\rm thick}\approx (63, 40)$~\kms~ and $(\sigma_{V_R},\sigma_{V_Z})_{\rm Halo}\approx (141, 94)$~\kms\ \citep{Soubiran03}.

\begin{figure*}
\centering

$\begin{array}{cc}
\includegraphics[width=0.45\textwidth, angle=0]{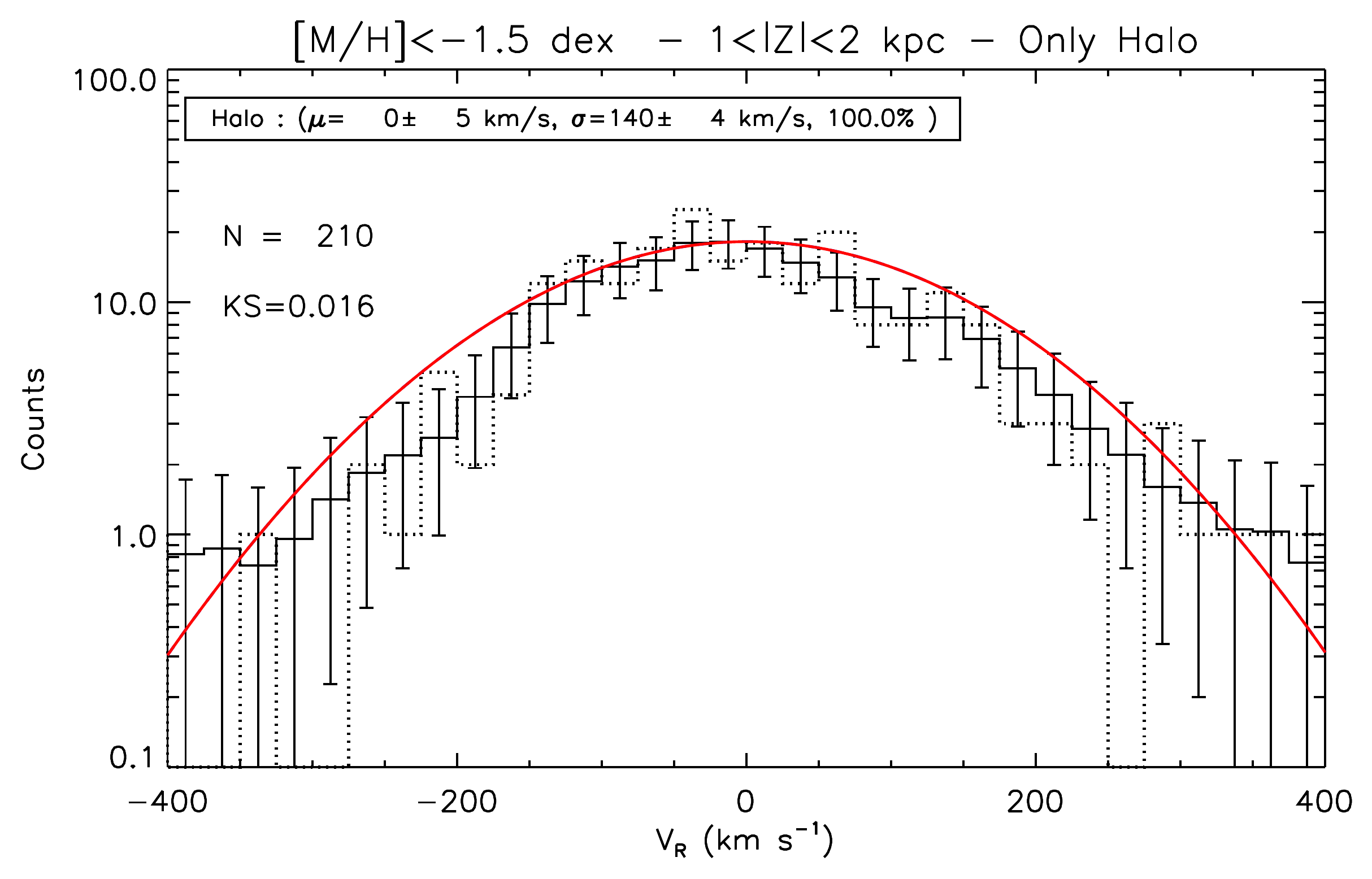} & \includegraphics[width=0.45\textwidth, angle=0]{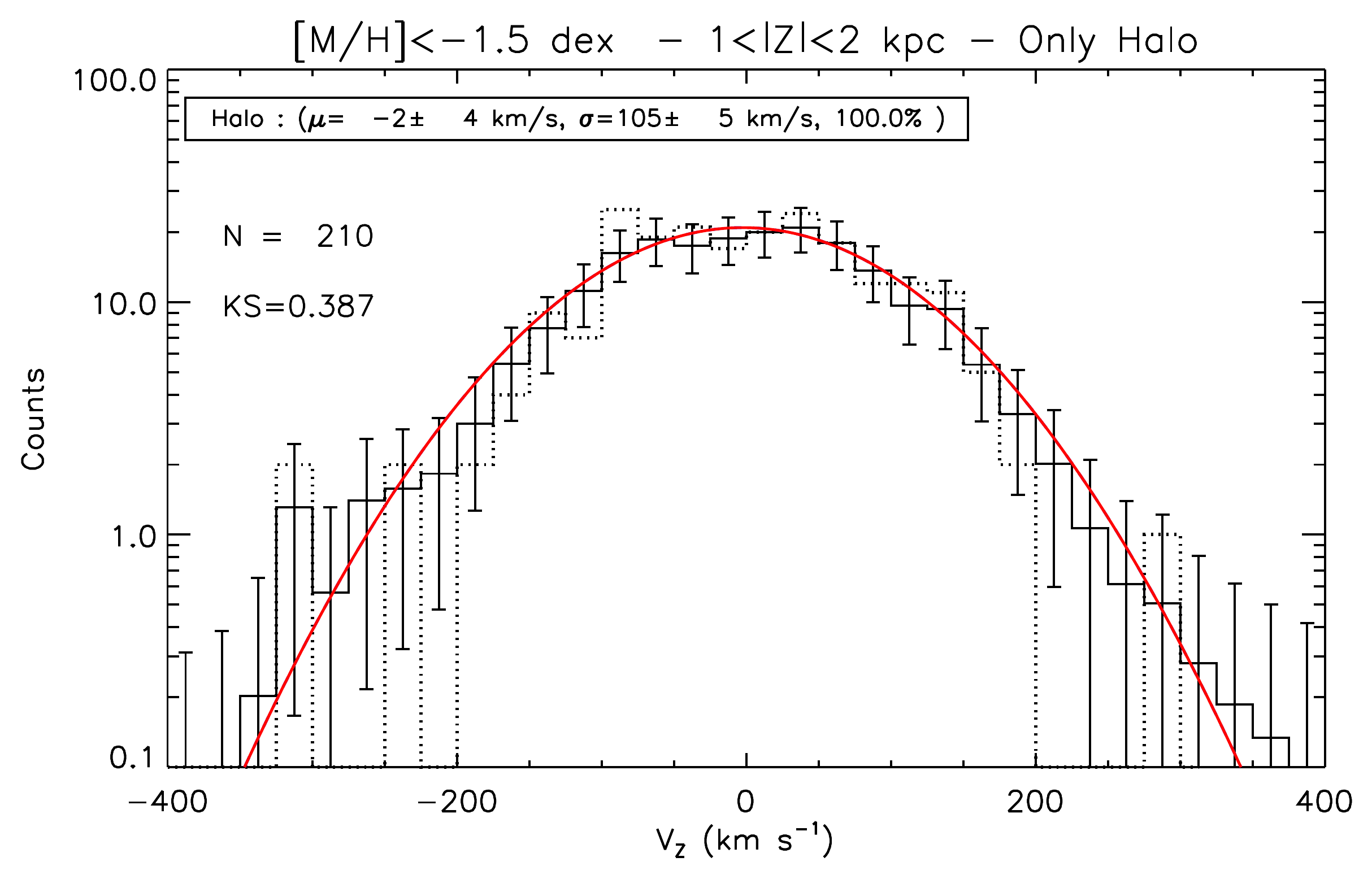}\\
\includegraphics[width=0.45\textwidth, angle=0]{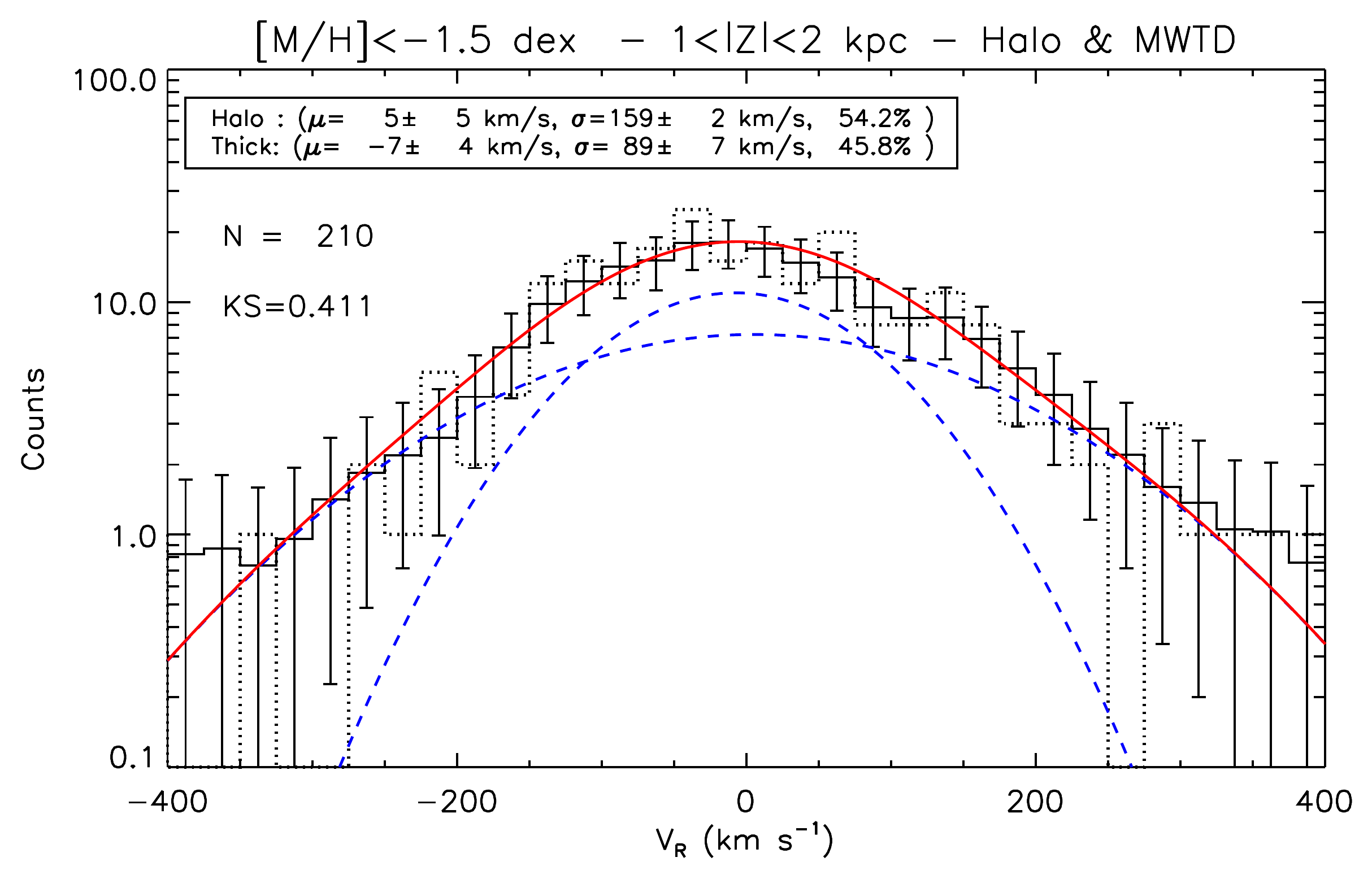}  & \includegraphics[width=0.45\textwidth, angle=0]{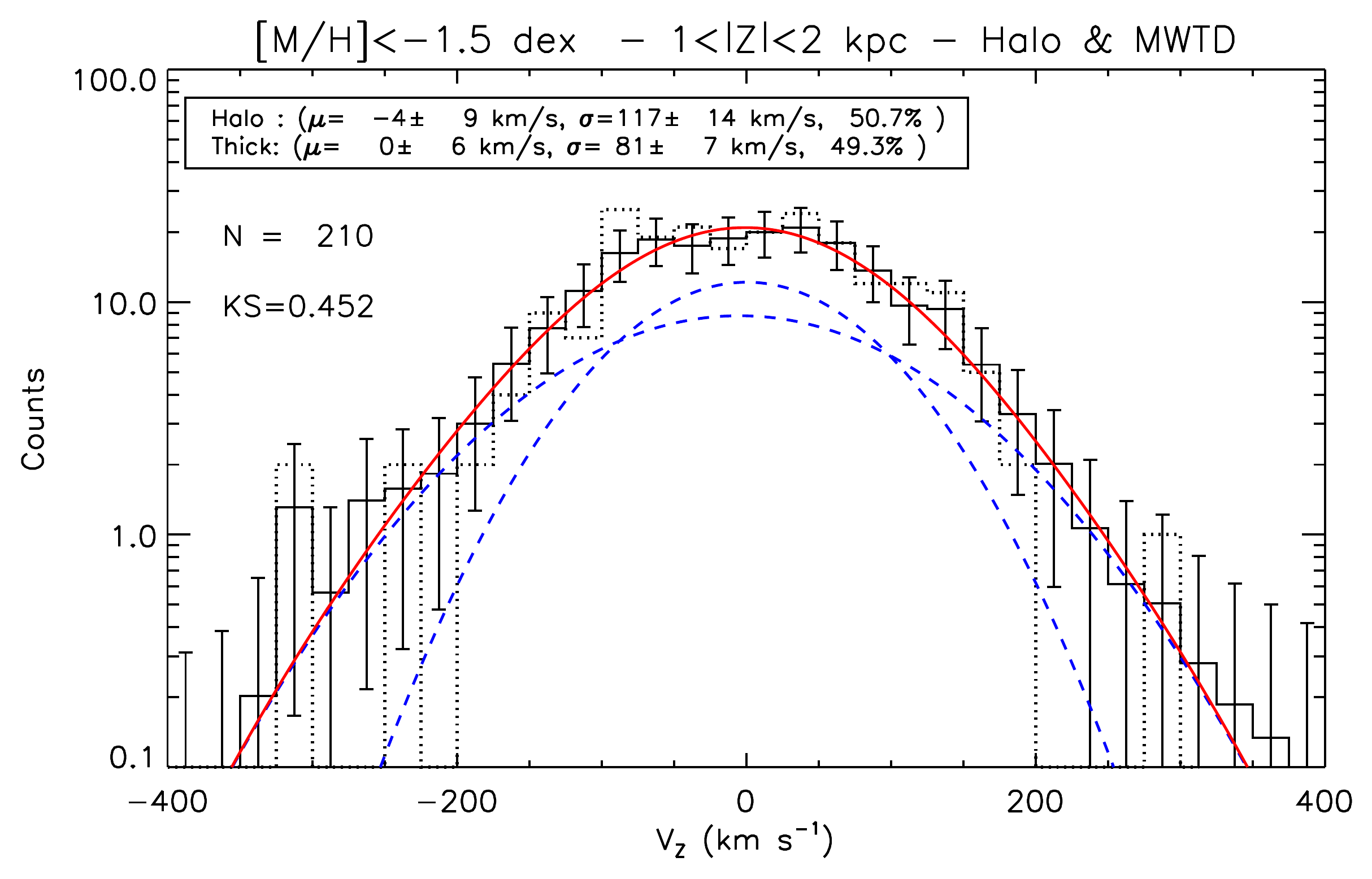} 
\end{array}$

\caption{Radial (left) and vertical (right) orbital velocity distributions for  all the stars with $\meta < -1.5$~dex being between 1 and 2~kpc far from the Galactic plane. The top panels show the fit when only one Galactic component, the halo, is used in order to fit the data, whereas the bottom panels consider two Gaussians (one for the thick disc and one for the halo).}
\label{fig:MP_Vz}
\end{figure*}

We adopt as our {\it a priori} radial and vertical velocity dispersions of the thick disc, the ones inferred in the previous section $(\sigma_{V_R}, \sigma_{V_Z} = 90, 60$~\kms). The vertical velocity dispersion for the stellar halo is also kept the same ($\sigma_{V_R}, \sigma_{V_Z} = 160, 130$~\kms). The fitting of the data show that a moderately good agreement can be found with two Gaussian components, with a ratio of thick disc stars over halo stars varying around $\sim 45-50\%$ (top panels of Fig.~\ref{fig:MP_Vz}). The ratio of thick disc to halo stars is close to the one found for the $\vphi$~distribution ($\sim 60 \%$).
We note nevertheless that the histogram of $V_Z$ is equally well fitted, with a single Gaussian having the characteristics of the halo (bottom panel of Fig.~\ref{fig:MP_Vz}), challenging the view of a MWTD if based only on the vertical velocity DF.
However, when the same fits are investigated for the radial component of the velocity, the KS-tests show that the two-component model fits much better the observations than a single Gaussian associated with the halo.  \\

\begin{figure}
\centering

$\begin{array}{c}
\includegraphics[width=1.\linewidth, angle=0]{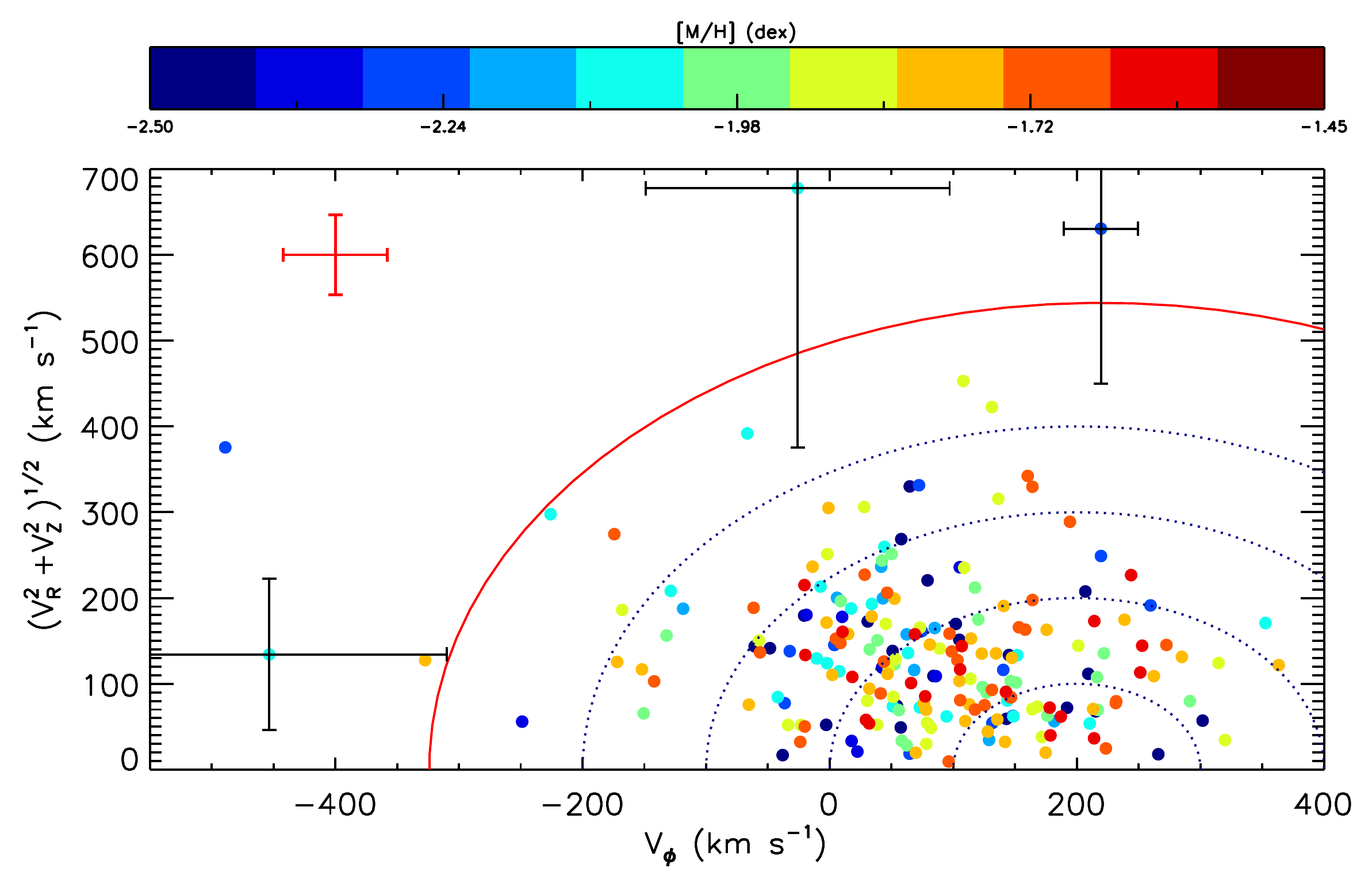}
\end{array}$

\caption{Toomre diagram for  all the stars with $\meta < -1.5$~dex and $1<|Z|<2$~kpc. The points are colour-coded according to their measured metallicity. All of the circles are concentric at $V_{LSR}=220$~\kms. The dotted circles represent total space velocities, $ V_{\rm tot}=\sqrt{V_R^2 + \vphi^2 + V_Z^2}$, in steps of 100~\kms. The local Galactic escape velocity speed \citep[$V_{esc}=544$~\kms,][]{Smith07} is plotted as a solid red circle. The mean error bar for the data is plotted in red in the upper left corner, whereas the black error bars are typical examples of the individual errors for the stars that are lying beyond that escape velocity. }
\label{fig:MP_Toomre}
\end{figure}

Figure~\ref{fig:MP_Toomre} shows the Toomre diagram for the considered metal-poor stars. We can see that there is a multitude of targets with intermediate angular momentum (thick disc-like kinematics, {\it i.e.} $\vphi > 120$~\kms) and unusually large meridional velocities ($\sqrt{V_R^2+V_Z^2} > 100$~\kms).
The mean error bar for the metal-poor sample is shown in red in the upper left corner of Fig.~\ref{fig:MP_Toomre}. It is easy to see that without the existence of a metal weak thick disc having the kinematic properties described above, the relatively small velocity uncertainties cannot explain only by themselves the high number of intermediate angular momentum stars.

On the other hand, we note that the few stars that have velocities larger than the local Galactic escape velocity \citep[$V_{esc}=544$~\kms,][]{Smith07}, the error bars are large enough to put them back within the Galactic binding energies (a few indicative individual error bars are plotted in Fig.~\ref{fig:MP_Toomre}). The characterisation of the stars having high $\vphi$~ will be discussed in a separate paper (Kordopatis et al., in prep.).


\subsection{On the plausibility of the existence of a metal-poor tail for the thick disc}

The previous sections showed that based on the star counts in the velocity space, similar relative normalisations of the Galactic components for the stars with  $\meta < -1.5$~dex at $1<|Z|<2$~kpc are observed, with proportions varying between 41--59\% depending on which distribution function is fitted. In this section we are investigating the plausibility of these numbers and infer down to which metallicity the thick disc is still detected.

Given the density numbers obtained in Table~\ref{table:measured_vels_sol_cyl_thick_disc} ($N_{thick} \sim 10 N_{halo}$), in order to reach a 50\% ratio at the lower metallicities, this implies that there is roughly 3\% of the thick disc that has metallicities lower than $-1.5$~dex.
Assuming the classical view of a canonical thick disc, modelled with a single Gaussian centred at $\meta\sim-0.5$~dex and a metallicity dispersion of 0.3~dex, the predicted star counts do not agree with the observations. Indeed, at $\meta \sim -1.6$~dex, we already are at roughly $3\sigma$ from the mean of the distribution, which implies that less than 0.1\% of the stars should belong to the thick disc.
In order to reach, as the observations suggest, such a large number of thick disc stars at low metallicities with only one Gaussian, the MDF should reach $\meta =-1.5$~dex at 2$\sigma$, implying that $(\sigma_\meta)_{thick}\approx 0.5$~dex.
We hence conclude that this model is not realistic given the histograms obtained for the entire RAVE sample (see, for example, upper plots of  Fig.~\ref{fig:Sol_cylinder} and Fig.~\ref{fig:Sol_cylinder_thickdisc}).\\

The existence of a metal-weak tail for the thick disc has already been previously discussed in the literature \citep{Norris85,Morrison90, Chiba00, Carollo10, Ruchti11, Kordopatis13a}.
By obtaining the metallicities by photometric means, and inferring the azimuthal velocities of a few hundred of stars observed close to the tangent point, where their radial velocity is predominated by the rotational velocity, \citet{Morrison90} claimed that in the abundance range $-1.6 < {\rm [Fe/H]} < -1.0$~dex, there are stars with disc kinematics, having the same density as the halo stars and a similar velocity lag as the canonical thick disc. According to these authors, the scale-length of the metal-weak thick disc is smaller than the one of the canonical thick disc. Our results (combined with the ones of Sect.~\ref{sect:scale_lenghts}) are consistent with this early study.

Ten years later, \citet{Chiba00}, observed 1203 stars in the Solar neighbourhood and  claimed that the metal-weak thick disc is visible down to metallicities [Fe/H] of $-2.2$~dex. They estimated that 10\% of the stars in the range $-2.2 < {\rm [Fe/H]} < -1.7$~dex belong to the thick disc, these metal-poor stars having a larger scale-length than their canonical thick disc.  As it will be shown in Sect.~\ref{sect:met_vel_correlation} (Fig.~\ref{fig:MP_Vphi_minus2}), hints of detection of the metal-poor thick disc down to metallicities of $-2$~dex are also obtained in our study.

Finally, with the advent of the large low-resolution spectroscopic survey of SEGUE, \citet{Carollo10} have claimed that the metal-weak thick disc has a comparable, though shorter, scale-length ($h_R\sim 2$~kpc) than the canonical thick disc ($h_R\sim 2.5$~kpc), and that it is detectable down to metallicities of $-1.8$~dex.\\

The first detailed attempt to chemically characterise the stars of the metal-weak thick disc has been done by \citet[][R10 hereafter]{Ruchti10}. The authors of that study obtained high-resolution spectra of 233 metal-poor stars ($\meta \lesssim -1$~dex) selected from the RAVE-DR3 catalogue \citep{Siebert11}. They found that for the metal-poor stars belonging to the thick disc (30\% of their targets), the $\alpha-$abundances are homogeneous and similar to the ones of the halo, {\it i.e.} with $[\alpha/{\rm Fe}]=+0.4$~dex. 

\begin{figure}
\centering
$\begin{array}{c}
\includegraphics[width=1.\linewidth, angle=0]{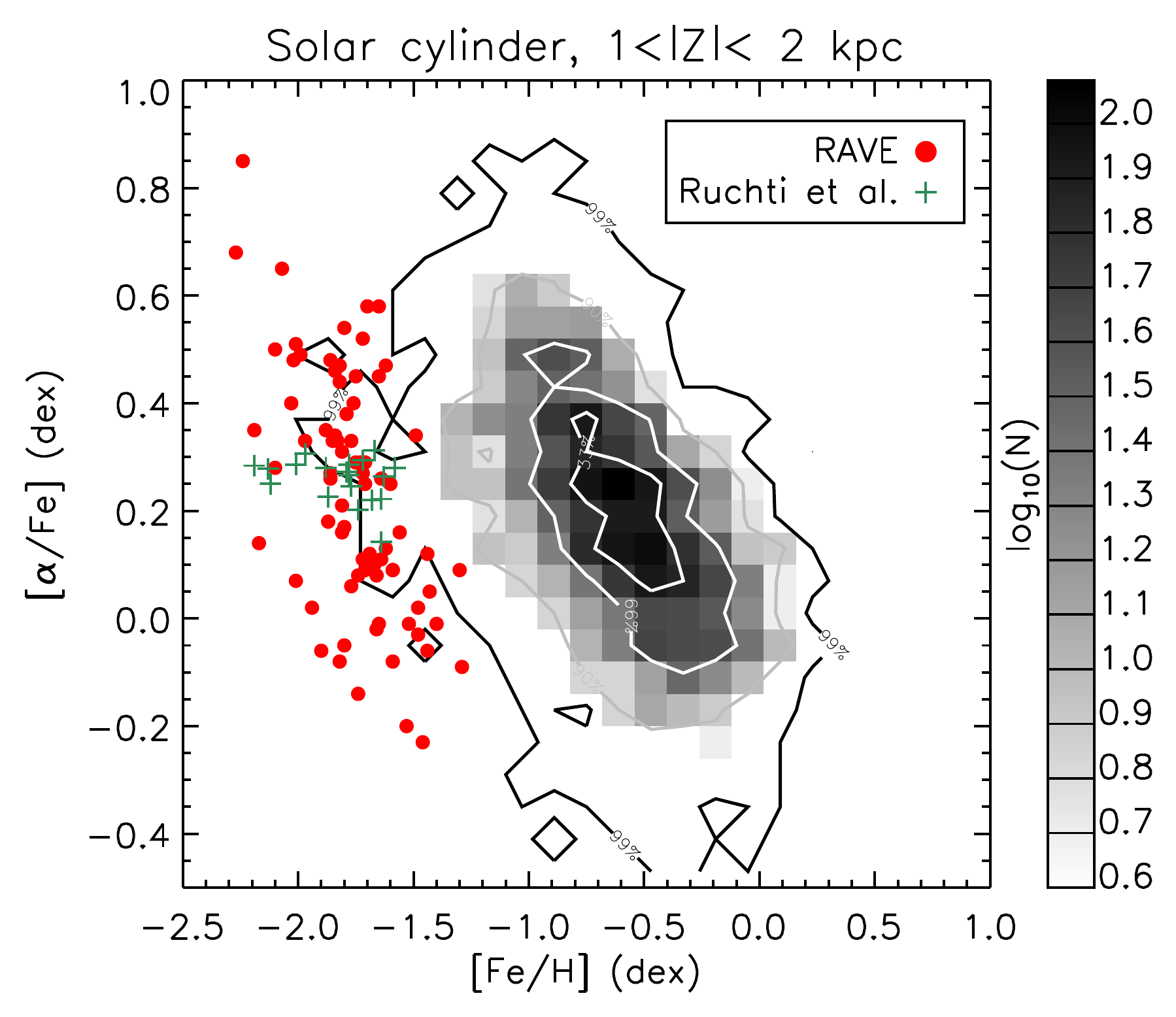}
\end{array}$
\caption{$[\alpha$/Fe] versus [Fe/H]  for the stars having $7.5 < R < 8.5$~kpc and $1<|Z|<2$~kpc (in grey scale) and for the potential stars of the metal-weak thick disc (red circles and green plus symbols). The contour-plot as well as the red circles are data coming from the chemical pipeline of RAVE, whereas the green "+" symbols are abundances of the stars for which high-resolution spectral analysis was performed by \citet{Ruchti10}. For the RAVE chemical pipeline, the $\alpha-$abundances are computed as the mean of ([Mg/Fe], [Si/Fe]. [Ti/Fe]), whereas for the \citet{Ruchti10} analysis, it is the mean of ([Mg/Fe],[Si/Fe], [Ca/Fe], [TiI/Fe], [TiII/Fe]). }
\label{fig:alpha_feh_mwtd}

\end{figure}

Our study uses the metallicities, the distances and the velocities obtained from different pipelines compared to the previous RAVE data-releases \citepalias[see] [for further details on the improvement and a comparison of the results obtained with each pipeline]{Kordopatis13b}. 
Among our selected stars with $\meta < -1.5$~dex and $1<|Z|<2$~kpc, 20 of them have individual elemental abundances measured by R10.  They have been represented in Fig.~\ref{fig:alpha_feh_mwtd} as green  $"+"$ symbols, together with the derived values obtained by the chemical pipeline of RAVE \citep[based on the much lower resolution spectra and the pipeline of ][]{Boeche11} in red filled circles.  
The big scatter in the RAVE individual abundance measurements, compared to the R10 ones, is due to the fact that at low metallicities equivalent-widths of spectral lines at the RAVE wavelength range and resolution are hard to measure, resulting in abundance estimates with large uncertainties \citep[see][and K13]{Boeche11}. However, despite this scatter, neither the RAVE abundances nor the R10 ones of the candidate   MWTD stars exhibit any particular chemical pattern differentiating them from what one would expect for the typical halo stars.

\subsection{Velocity-metallicity correlation}
\label{sect:met_vel_correlation}
A correlation between the metallicity and the azimuthal velocity is the natural outcome of the epicyclic motion of the stars. This is  due to the conservation of the angular momentum during the epicycle excursions (stars from the outer disc increase their azimuthal velocity whereas stars from the inner disc decrease it) and to the fact that inner disc stars are more metal-rich than outer disc ones \citep[see ][for values of radial metallicity gradients]{Rudolph06,Pedicelli09, Andreuzzi11, Gazzano13}.   
On the other hand, radial migration mechanisms change the stellar orbits without conservation of the angular momentum. As a consequence, depending on the efficiency of the radial migration mechanisms, the amplitude of this correlation will change \citep[see for example,][]{Haywood08}. 
In the particular case of the formation of the thick disc through such internal processes, the amplitude of this correlation is expected to be, if not null, at least small \citep{Loebman11}. The main argument for that characteristic is that at a given small radial range (like for example at the Solar cylinder), stars of different metallicities as old as the ones of the thick disc ($\sim 10$~Gyr)  should have been fully mixed, thus erasing that kind of correlation signature between rotation and metallicity.

Without assigning probabilistically each target to the halo or the thick disc, our analysis can still give an estimation of this correlation based on the results of the previous sections. Considering that the scoped metal-weak thick disc has $\meta \sim -1.6$~dex, at a  mean azimuthal velocity of $\vphi \sim 123$~\kms\ (Sect.~\ref{sec:MWTD_vphi}), then this implies that $\partial \vphi/\partial \meta \approx 50$~\kms~dex$^{-1}$. This estimation  is in very good agreement with the correlation value that can be found in the literature \citep{Spagna10,Kordopatis11b, Lee11}, measured either on kinematically or space-selected targets of the canonical thick disc.

\begin{figure}
\centering
$\begin{array}{c}
\includegraphics[width=1.\linewidth, angle=0]{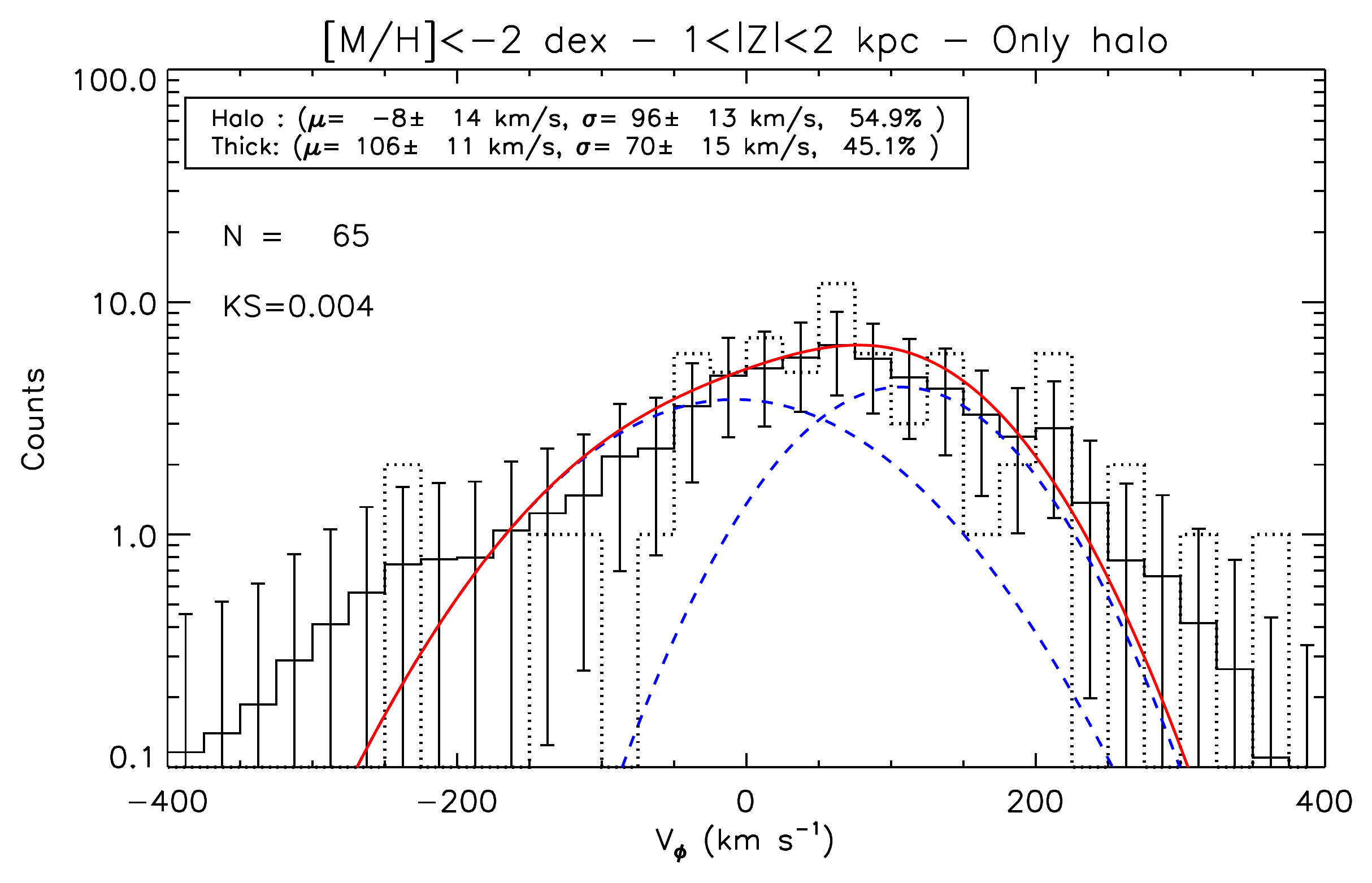}
\end{array}$
\caption{Azimuthal velocity distribution for the RAVE stars with $\meta < -2$~dex, and $1<|Z|<2$~kpc. In spite of the low number statistics, hints of the detection of the  metal-weak thick disc are still obtained.}
\label{fig:MP_Vphi_minus2}
\end{figure}

We further investigated towards the lower metallicities to test if this estimated correlation is confirmed. In this application we select only the stars with $\meta < -2$~dex. To the 65 stars that have been selected that way we repeat the same Maximum Likelihood fit as described in the previous sections. As it can be noticed on Fig.~\ref{fig:MP_Vphi_minus2}, the metal-weak thick disc  can still be observed, in a proportion of 45\%.  The azimuthal velocity has also decreased. The value expected from the correlation $\partial \vphi/\partial \meta \approx 50$~\kms~dex$^{-1}$ is met, hence indicating a MWTD peaking at $\vphi \approx 110$~\kms.

{  
\subsection{Estimations of the true velocity dispersions and the spatial distribution of the metal-weak thick disc}
\label{sect:scale_lenghts}
The previously estimated velocity dispersions have not been corrected from the error measurements, due to the challenging task of estimating the mean error of the stars belonging to each component (see Sect.~\ref{sect:Solar_cylinder}). 
Nevertheless, an estimation of the error-free velocity dispersions can be inferred for the thin disc, thick disc and metal weak thick disc considering the samples where they are the dominant population, and by subtracting according to Eq.~\ref{eqn:dispersion_correction} the measurement errors. 

The most accurate parameters to be used concerning the thin disc are obtained by fitting the Solar cylinder sample (thin disc dominated, Table~\ref{table:measured_vels_sol_cyl}). The ones concerning the thick disc are obtained by fitting the Solar cylinder sample at $1<|Z|<2$~kpc (thick disc dominated, Table~\ref{table:measured_vels_sol_cyl_thick_disc}). As far as the metal-weak thick disc is concerned, the fits to be used come from the sample with $\meta<-1.5$~dex and $1<|Z|<2$~kpc ($\sim 50\%$ of MWTD, Table~\ref{table:measured_vels_MW_thick_disc}).
Since for these samples the concerned populations are predominant, we use the averaged error values of all the stars in each sub-sample to correct the velocity dispersions. The derived mean errors that are used are: 
\begin{itemize}
\item
$(eV_R,e\vphi,eV_z)_{thin}=(10.4,8.1,7.4)$~\kms

\item
$(eV_R,e\vphi,eV_z)_{thick}=(35.2,28.6,21.6)$~\kms

\item
$(eV_R,e\vphi,eV_z)_{MWTD}=(41.7, 48.0, 37.5)$~\kms.
\end{itemize}

The resulting velocity dispersions are shown in Table~\ref{table:corrected_vels}.
As far as the results for the thin disc are concerned, they are in good agreement with \citet{Pasetto12b} who also used RAVE data, or the ones of \citet{Nordstrom04} from the Geneva-Copenhagen Survey, though colder than the ones of \citet{Soubiran03}. 
For the thick disc, our derived results are hotter than the ones presented in \citet{Pasetto12} and \citet{Carollo10} 
but are in relative agreement with \citet{Casetti-Dinescu11}. 
Finally, the estimated values for the metal weak thick disc are in disagreement with \cite{Carollo10}  who estimated $(\sigma_{V_R}, \sigma_{\phi},\sigma_{V_Z})=(59,40,44)$~\kms, but in agreement with the lagging thick disc of \citet{Gilmore02} who suggested $(\sigma_{V_R}, \sigma_{\phi},\sigma_{V_Z})=(63,70,60)$~\kms\ \citep[see also][]{Ruchti11}.

\begin{table*}
\centering
\caption{Mean velocities and dispersions corrected from error measurements for the disc components of the Milky Way}
\begin{tabular}{c c c c c c c }
\hline\hline
Galactic component & $V_R$ & $\vphi$ & $V_Z$ & $\sigma_{V_R}$& $\sigma_{V_\phi}$  & $\sigma_{V_Z}$  \\
                & \kms         & \kms             & \kms  & \kms         & \kms             & \kms \\
\hline
Thin disc &  \vrdmcyl &  \vphidmcyl  &  \vzdmcyl  & $26 \pm 1$  & $16 \pm 1$ &  $15 \pm 1$  \\
Thick disc &  \vrdeTD  &  \vphideTD &  \vzdeTD & $71 \pm 4$ & $44 \pm 1$ & $45 \pm 2$   \\
MWTD &  \vrdeMWTD & \vphideMWTD & \vzdeMWTD & $67\pm 7$ & $61 \pm 12$  & $61 \pm 6$ \\
\hline
\end{tabular}
\label{table:corrected_vels}
\end{table*}

Interestingly, Table~\ref{table:corrected_vels} also indicates that  the canonical thick disc velocity dispersions and the ones of the metal weak thick disc are similar, the lag of the latter being also larger. 
Qualitatively, this implies that the scale-length of the MWTD is smaller than the one of the canonical thick disc\footnote{No quantitative results on the relative spatial extension of the thick disc and MWTD are given, due to the difficulty on including pertinent parameters in the Jeans equations, such as the effective radius at which the measurements are being made  (see Fig.~\ref{fig:MP_positions}).}.
This statement is in agreement with \citet{Morrison90} and \citet{Carollo10},  but contradicts the ones of \cite{Chiba00} and \citet*{Bovy12}. Indeed, using a similar mixture model as \citet{Nemec91,Nemec93}, \citet{Bovy12b} gave for a combination of mono-abundance populations from SEGUE estimations on the scale-lengths and scale-heights and suggested that what is called in this work the {\it metal-poor thick disc} ([Fe/H]$\approx -1.3$~dex, [$\alpha$/Fe]$\approx +0.4$~dex) should have a similar scale-length to the ``thick" disc at any other metallicity (though with different scale-heights) of roughly 2~kpc.


 }

\section{Discussion: implications on the formation mechanisms of the thick disc}
\label{sect:discussion}
The non-Gaussianities of the metallicity and the velocity distributions are not a surprise since we expect them to be naturally skewed, in particular in the velocity space. However, the metallicities down to which we can still identify the thick disc and the correlation between [M/H] and $\vphi$ put strong constraints on the formation mechanisms of that structure \citep[e.g.:][]{Calura12}.
For the sake of completeness, we briefly discuss below 
 these constraints with regards to the four main mechanisms regularly evoked to form the thick disc: radial migration, several minor mergers, and a major stellar or gas-rich accretion, bearing in mind that a combination of two or more of the below mechanisms is most plausible to be at the origin of the thick disc \citep[e.g.:][]{Brook12, Liu12, Minchev12, Boeche13}.

\subsection{Radial migration}

The mechanisms involved for the radial migration of the stars, as described in \cite{Sellwood02} imply that the stars in co-rotation with the transient spiral arms \citep[but also the Galactic bar, see ][]{Minchev10,Minchev11} can migrate and change their radial orbit without any dramatic change in their orbital random energy, though changing their angular momentum. Radial migration predicts that the most metal-poor disc stars at the Solar neighbourhood have migrated from the outskirts of the Galaxy.

The migration happening predominantly from the inner Galaxy towards the outer radii, the models of \cite{Schonrich09b, Loebman11} could presumably  explain why we observe such a small number of metal-poor stars. 
Nevertheless, it should be noted that stars migrating outwards from the inner disc provide minimal contribution to the vertical velocity dispersion (and thus disc thickness) at a given outer radius. Indeed, the assumption that stars migrating outwards preserve their vertical energy, is not true, since in fact it drops exponentially because the conserved quantity is in fact the vertical action \citep*{Solway12}. This by itself already rules out radial migration as the source of forming a thick disc \citep{Minchev12c}. 
In addition, in that frame, it cannot be explained how the metal-poor stars having migrated from outwards could have reached high distances above the Galactic plane ($1<|Z|<2$~kpc), in particular because they migrate towards regions where the Galactic potential is higher.

One could also argue that these metal-poor stars come from the inner radii of the Galaxy. In that case, it is expected that these metal-poor stars would have spread in the Galaxy through the churning processes, and hence this could explain their relatively low density. There are no age issues with this scenario, since the inner part of the Galaxy had a very active star formation rate, and hence could have had a very broad metallicity distribution function.
On the other hand, if this scenario is true, then one should expect to find many metal-rich stars in the Solar neighbourhood, at the distances where the thick disc dominates, since they should have migrated in a similar manner.
The metal-rich end of RAVE-DR4 is for the moment poorly constrained, so further investigation must be done in the future.


\subsection{Dynamical heating due to minor mergers}

This scenario implies that the pre-existent thin disc has been dynamically heated after successive accretions of smaller stellar systems \citep{Toth92,Quinn93,Kazantzidis08,Villalobos08,Moster10, DiMatteo11}. In that case, the metal-poor thick disc would most probably be the extra-galactic component, whereas the 'canonical' thick disc would mainly be composed from stars born {\it in situ} heated from the accretion  of the merging dwarfs on disc-like orbit \citep*{Gilmore02}. The main problem with this scenario is that there is no obvious reason why these two populations should follow the metallicity-velocity correlation that is observed, unless the minor mergers did not retain any kinematic memory.

\subsection{Gas-rich accretion}

This scenario involves an accretion-driven star formation, where in particular the stars of the thick disc are mainly born {\it in situ}, though from extra-galactic gas 
{\citep[e.g.][]{Brook04, Brook12}}.
 This star formation event  must have been quite short, the main observational constraints being the high [$\alpha$/Fe] that are measured for the thick disc. 
 In that scenario, it is possible to have  the extra-galactic stars and the {\it in-situ} born thick disc stars with the metallicity-velocity correlation characterising the thick disc stars.

\subsection{Satellite accretion}

In this scenario, most of the thick disc stars are extra-galactic, accreted from a satellite galaxy. \citet{Abadi03} have shown that a disc can survive a fairly massive accretion event, if the satellite comes in on a relatively prograde and co-planary orbits (up to original inclinations of 20-30 deg, as tides and dynamical friction pull the satellite into the plane before accretion).
 This scenario can explain rather well the metal-weak stars, as they would have formed after a process of self-enrichment, like it is observed for the dwarf spheroidal galaxies orbiting the Milky Way. Nevertheless, such a scenario is challenging, since in order for a dwarf galaxy to reach the typical metallicities of the thick disc of $\sim -0.5$~dex 10~Gyr ago (which is the mean age of the thick disc stars) a galaxy of considerably high mass is required \citep{Wyse08}, which would have destroyed the pre-existent thin disc \citep[though see][for a discussion on the effect of a gaseous disc during the accretion events]{Moster10}.


\subsection{Identification with moving groups or streams?}
Here we discuss the possibility that the detected intermediate angular momentum and low metallicity population that we associated with the metal-weak thick disc belongs in fact to one of the streams or moving groups in the Galaxy that have similar kinematic and/or chemical properties as the ones described in this study. 

As first noted by \cite{Carollo10}, the Monoceros stream \citep{Yanny03} presents similar properties in both the kinematic and metallicity as the MWTD. It is still a matter of debate if the Monoceros stream is the remnant of a dwarf galaxy cannibalised  by the Milky Way, or rather a signature of a more complex thick disc  \citep{Lopez12}. In any case, the suggested stream is mainly observed in the second and third quadrant of the Milky Way, {\it i.e.} towards the anti-centre. As it can be seen in Fig.~\ref{fig:MP_positions}, this is not the case for the considered RAVE sample, since the selected metal-poor stars in the survey are located towards the inner Galaxy.
Further investigations are needed, in terms of orbit determination and individual abundances determinations ($r-$ and $s-$ elements), in order to potentially associate the MWTD we identified to the Monoceros group. 
If these investigations were to suggest that 
the stars identified as MWTD stars  were members of the Monoceros over density, this result would promote the argument of a Galactic origin of this stream.

Similarly to the Monoceros stream, the mean azimuthal velocity of the Arcturus moving group \citep[e.g.:][]{Navarro04,Ramya12}, is roughly $\sim 120$~\kms. However, it is unlikely that what we identified as the MWTD belongs to the Arcturus moving group, as its metallicity is low (compared to the mean value of $-0.6$~dex for the Arcturus group), homogeneously spread in the Galaxy, with a radial and vertical velocity centred at 0~\kms, and a correlation between the kinematics and the metallicity.

\section{Conclusions}
\label{sect:conclusions}
Using the latest data release of RAVE (DR4) we fitted the chemical and velocity space of the data using a simple Galactic model composed by an old thin disc, a canonical thick disc and a stellar halo. 
We have shown that this simple model describes relatively well the overall distributions, though an {\it extra} component is required in order to appropriately  fit the velocity distributions of stars with metallicities lower than $\meta < -1.5$~dex.  
 We found that this population is entirely consistent with the thick disc in terms of velocities, since it follows the rotation-metallicity correlation established by other, independent surveys for the thick disc stars based on kinematically or space selected samples.  

This Metal Weak Thick Disc represents at least 3\% of the canonical thick disc, reaches metallicities down to $-2$~dex, and is characterised by a shorter scale-length than the canonical thick  disc. 
Finally, the implications of such a metal-weak tail for the thick disc have been discussed in the frame of the thick disc formation scenarios, challenging in particular the mechanisms of radial migration.

\section*{Acknowledgments}
We thank A.~Helmi, J.~Binney and the anonymous referee for valuable comments on the
manuscript. 
 Funding for RAVE has been provided by: the Australian
Astronomical Observatory; the Leibniz-Institut fuer Astrophysik Potsdam (AIP); the Australian National University;
the Australian Research Council; the French National Research Agency; the German Research Foundation (SPP 1177
and SFB 881); the European Research Council (ERC-StG
240271 Galactica); the Instituto Nazionale di Astrofisica at
Padova; The Johns Hopkins University; the National Science Foundation of the USA (AST-0908326); the W. M.~Keck foundation; the Macquarie University; the Netherlands Research School for Astronomy; the Natural Sciences
and Engineering Research Council of Canada; the Slovenian Research Agency; the Swiss National Science Foundation; the Science \& Technology Facilities Council of the UK;
Opticon; Strasbourg Observatory; and the Universities of
Groningen, Heidelberg and Sydney. The RAVE web site is
at \url{http://www.rave-survey.org}.

\bibliographystyle{mn2e}

\bibliography{Metal_poor}

\end{document}